\pdfoutput=1

\documentclass[11pt]{article}

\usepackage[final]{acl}

\usepackage{times}
\usepackage{latexsym}

\usepackage[T1]{fontenc}

\usepackage[utf8]{inputenc}

\usepackage{microtype}

\usepackage{inconsolata}

\usepackage{graphicx}
\usepackage{multicol}
\usepackage{enumitem}
\usepackage{CJKutf8}
\usepackage{algorithm}
\usepackage{amsmath}
\usepackage{amsthm}
\usepackage{booktabs}
\usepackage{tabularx}
\usepackage{colortbl}
\usepackage{amssymb}
\usepackage{subfigure}
\usepackage{multirow}
\usepackage{arydshln}
\usepackage{algpseudocode}  

\usepackage{ulem} 
\usepackage{tikz} 

\usepackage{makecell}

\definecolor{mlb}{RGB}{173,216,230}  
\definecolor{mlo}{RGB}{255,223,186}  

\title{Fine-grained Video Dubbing Duration Alignment with Segment Supervised Preference Optimization}

\author{
 \textbf{Chaoqun Cui\textsuperscript{1}},
 \textbf{Liangbin Huang\textsuperscript{1,2}},
 \textbf{Shijing Wang\textsuperscript{3}},
 \textbf{Zhe Tong\textsuperscript{1}}
 \textbf{Zhaolong Huang\textsuperscript{1}},
\\
 \textbf{Xiao Zeng\textsuperscript{1}},
 \textbf{Xiaofeng Liu\textsuperscript{2}\thanks{Corresponding author.}}
\\
\\
\textsuperscript{1}Alibaba Digital Media and Entertainment Group
\\
\textsuperscript{2}School of Software Engineering, Huazhong University of Science and Technology
\\
\textsuperscript{3}Beijing Key Laboratory of Traffic Data Mining and Embodied Intelligence,\\Beijing Jiaotong University, Beijing 100044, China
\\
 \small{
   \textbf{Correspondence:} \href{cuichaoqun.ccq@alibaba-inc.com}{cuichaoqun.ccq@alibaba-inc.com}, \href{liuxf@hust.edu.cn}{liuxf@hust.edu.cn}
 }
}

\begin{document}
\maketitle
\begin{abstract}

Video dubbing aims to translate original speech in visual media programs from the source language to the target language, relying on neural machine translation and text-to-speech technologies. Due to varying information densities across languages, target speech often mismatches the source speech duration, causing audio-video synchronization issues that significantly impact viewer experience. In this study, we approach duration alignment in LLM-based video dubbing machine translation as a preference optimization problem. We propose the Segment Supervised Preference Optimization (SSPO) method, which employs a segment-wise sampling strategy and fine-grained loss to mitigate duration mismatches between source and target lines. Experimental results demonstrate that SSPO achieves superior performance in duration alignment tasks.

\end{abstract}

\section{Introduction}

Video dubbing involves translating the original speech from a source language to a target language in visual media programs, relying on machine learning speech language processing techniques. Typically, video dubbing systems are not end-to-end but consist of three cascaded sub-tasks \cite{2,1}, namely Automatic Speech Recognition (ASR) \cite{3-1,3-2}, Neural Machine Translation (NMT) \cite{5-2,5-1}, and Text-to-Speech (TTS) \cite{4-3,4-2,4-4,4-1}. ASR converts the original speech into text. When subtitles are available or can be obtained through Optical Character Recognition (OCR) \cite{6-1,6-2}, ASR can be bypassed. NMT is used to translate the source language text to the target language, after which TTS synthesizes the translated text into speech in the target language.

In video dubbing systems, maintaining strict isochronous constraints between the original source speech and the synthesized target speech in terms of speech duration is crucial for ensuring synchronization with the original video footage, which is vital for preserving an immersive experience for the audience \cite{1}. However, due to varying information densities across different languages, translating from one language to another often results in a duration mismatch between the source speech and the target speech \cite{mismatch1,mismatch2,10}. For instance, when translating from Chinese, a high information density language, to lower information density languages such as English or Thai, the resulting translations frequently exceed the timing notes of the original subtitles (see Table~\ref{tab:subtitle} for examples of subtitle format.), significantly impacting the audience's viewing experience. If relying solely on TTS to adjust the pause and duration of words, the vast differences in information density between languages necessitate that TTS adjusts the speaking rate of each word within a wide range to match the total speech duration. This can severely impact the fluency and naturalness of the synthesized speech, leading to a dissonance in the speaking rates between adjacent lines \cite{1}. Consequently, Duration Alignment (DA) is a significant challenge that must be addressed during NMT.

Recently, Large Language Models (LLMs) have been widely applied in NMT, bringing significant improvements to translation tasks \cite{21-1,21-2,21-3,add1}. LLMs have also been applied to video dubbing NMT. However, relying solely on Prompt Engineering (PE) and Supervised Fine-tuning (SFT) on human-translated subtitles does not handle DA well. This is primarily because LLMs lack direct awareness of speech duration for the text, and the available human-translated subtitles used for SFT typically focus on the text itself rather than the speech duration of the lines. For DA, although generating a translation that is a few words shorter or longer may seem like a simple task, it actually requires good control over the target language. As illustrated in Table~\ref{tab:casestudy}, LLMs must implicitly adopt strategies such as choosing more concise phrasing, using different verb tenses, avoiding redundant adverbs and adjectives \cite{10}, while also maintaining translation accuracy and fluency to ensure that the translation quality does not deteriorate after DA.

In this study, we consider the DA task as a special preference optimization problem, termed as a localized multi-segment preference optimization problem. This framework addresses the situation where the output of LLMs consists of multiple semantically interconnected segments. The preference metric is evaluated for individual segments rather than the entire outcome (segment supervised), requiring segment-wise alignment across each segment. To tackle this issue, we propose the \textbf{S}egment \textbf{S}upervised \textbf{P}reference \textbf{O}ptimization (SSPO) method. SSPO effectively controls the duration of translations for each line at a fine-grained level while ensuring accuracy and fluency in handling DA tasks.

In summary, this study contributes as follows:
\begin{itemize}
\item We define duration consistency metrics and established an evaluation framework for DA.
\item We propose the SSPO method, which formulates DA as preference optimization problem.
\item We elucidate the theoretical foundation of SSPO and formalize the localized multi-segment preference optimization task.
\item Experimental results demonstrate the effectiveness and robustness of SSPO.
\end{itemize}

\section{Related Work}

\subsection{Duration Controllable Generation}

Previous research on text length control includes: 1) using reward functions or models incorporating length information to guide decoding \cite{12,11}; 2) modifying model embeddings to inject length information \cite{10,13}; and 3) interfering with training using length prediction metrics or models \cite{14,1,pep}. However, these methods primarily target traditional sequence-to-sequence models and are unsuitable for LLMs. This is because LLMs are highly optimized through large-scale pre-training, and modifying their embeddings or interfering with training would significantly degrade overall performance \cite{embedding1,embedding2}. Moreover, these methods aim to generate shorter texts, whereas DA seeks to produce translations with consistent durations to the original. Additionally, subtitle dialogues are short texts with strong contextual relationships \cite{avt1,avt2,ragcl,adgscl}. In summary, unlike length control tasks aiming for overall shorter texts, DA's objective is to \textit{generate fine-grained translations for each line in LLMs' responses, ensuring consistent (not necessarily shorter) durations for individual lines rather than the entire response}.

\subsection{Language Model Preference Optimization}

Reinforcement learning offers an effective solution for aligning LLMs with human values and controlling text generation \cite{25,26,27,29,30,31}. Reinforcement learning from human feedback (RLHF) framework has been developed, based on human feedback reward models \cite{33,34,35,36,28,16-2}. Despite RLHF's effectiveness, its complexity, instability, and hyperparameter sensitivity remain unresolved \cite{38,37}. Recently proposed DPO \cite{17} simplifies the RLHF framework by eliminating the need for explicit reward modeling or reinforcement learning processes, thus avoiding dependence on reward models. Several variants have emerged, including SimPO, KTO, and IPO \cite{18,19,20}. However, these methods still face limitations such as coarse granularity and gradient dilution when addressing localized preference alignment tasks like DA.

\section{Preliminaries}

\subsection{Notations}

DA of video dubbing is essentially a task of controllable text generation (CTG) \cite{15} task, which requires the LLM's output to: 1) strictly conform to the corresponding format, so that the translation of each line can be matched with the original; 2) maintain the duration of each line's translation as consistent as possible with the original. Specifically, we leverage human-translated subtitles to perform SFT on an "off-the-shelf" LLM \cite{22,23,24}, and then perform DA on the SFT model. During SFT, the LLM's input prompt $x$ includes the instruction, a terminology translation table, and a set of $n$ source lines $s_{1},s_{2},\dots,s_{n}$ to be translated. The LLM's response $y$ includes the original and translated lines $s_{1},t_{1},s_{2},t_{2},\dots,s_{n},t_{n}$, where the original lines are output to avoid mismatches caused by model omissions or line merging. Although this generates more output tokens, it is crucial for ensuring the accuracy and correspondence of the translation. After DA, the output format of the LLM needs to be consistent with the SFT model, and it is necessary to maintain the duration consistency between $s_{i}$ and $t_{i}$. In our experiments, we consistently set $n=35$. Examples of the model's prompt and response are shown in Table~\ref{tab:prompt} and Table~\ref{tab:response} in Appendix~\ref{sec:io}.

\subsection{Preference Metric}
\label{sec:metric}

In our experiments, we utilize Microsoft Edge's online TTS service \texttt{edge-tts}\footnote{\url{https://github.com/rany2/edge-tts}} to obtain the duration of dialogue lines. It can be replaced with any TTS component. We use \texttt{edge-tts} to synthesize speech for both the original text $s_{i}$ and the translated text $t_{i}$ of each dialogue line, and then acquire their respective durations $\text{Dur}(s_{i})$ and $\text{Dur}(t_{i})$. Subsequently, we employ the following metric to measure the duration consistency between $s_{i}$ and $t_{i}$:

\begin{equation}
\resizebox{0.48\textwidth}{!}{$
\begin{aligned}
\mathcal{P}(s_i,t_i)=&\text{exp}(\max(0,\text{Dur}(t_i)-\text{Dur}(s_i)))\\&+\max(0,\text{Dur}(s_i)-\text{Dur}(t_i))-1.
\end{aligned}
$}
\end{equation}

\begin{figure}[!h]
  \centering
  \includegraphics[width=0.35\textwidth]{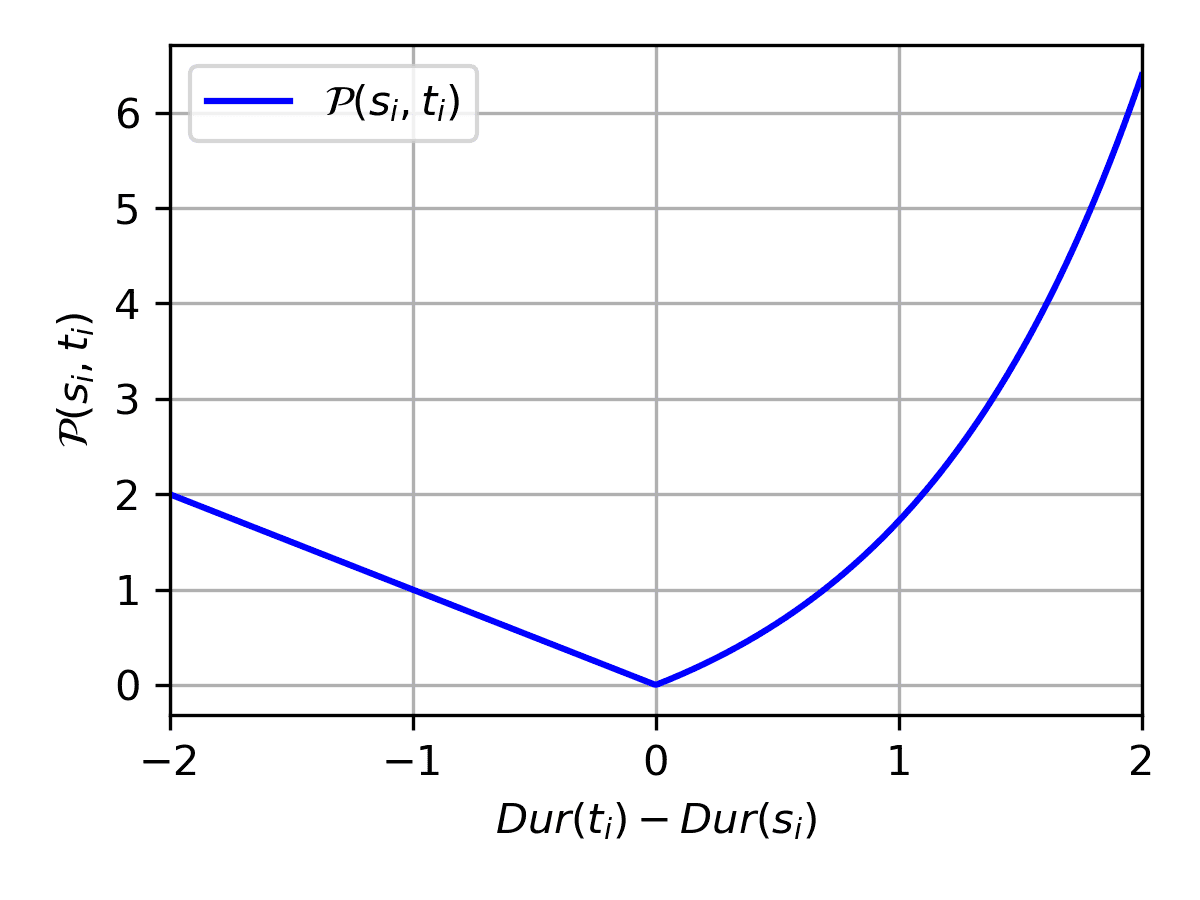}
  \caption{The function graph of $\mathcal{P}(s_i,t_i)$.}
  \label{fig:functionp}
\end{figure}

$\mathcal{P}(s_i,t_i)$ represents the penalty imposed when $\text{Dur}(s_{i})$ and $\text{Dur}(t_{i})$ are inconsistent, as shown in Figure~\ref{fig:functionp}. When $\text{Dur}(t_{i})>\text{Dur}(s_{i})$, $\mathcal{P}(s_i,t_i)$ is an exponential term, and when $\text{Dur}(t_{i})<\text{Dur}(s_{i})$, $\mathcal{P}(s_i,t_i)$ is a linear term. This design is based on the consideration that for video dubbing, longer translation duration is less acceptable than shorter one, as they may cause the translated subtitles to exceed the timing notes range of the original subtitles. The lower $\mathcal{P}(s_i,t_i)$ is, the higher the duration consistency between $s_{i}$ and $t_{i}$, and vice versa.

To evaluate the translation quality during the subsequent sampling process, we employ two widely used reference-free translation assessment models: \texttt{Unbabel/wmt23-cometkiwi-da-xxl} (denoted as KIWI-XXL) \cite{kiwixxl} and \texttt{Unbabel/XCOMET-XXL} (denoted as XCOMET) \cite{xcomet}. Both models have 10B parameters and demonstrate high correlation with human judgments.

\section{Method}

\subsection{Overall Framework}

Although we have defined a quantitative duration consistency metric $\mathcal{P}(s_i,t_i)$ in Section~\ref{sec:metric}, we are still unable to design a differentiable loss function to directly optimize the SFT model for DA. This is primarily because LLMs do not directly generate text, but rather predict the probability of token generation \cite{gpt}. Consequently, utilizing a metric like $\mathcal{P}(s_i,t_i)$ to optimize the LLM directly through gradient descent is infeasible. In light of this, we approach DA as a preference optimization problem. Within the preference optimization framework, we can leverage the $\mathcal{P}(s_i,t_i)$ metric to guide the generation probabilities of the LLM, thereby optimizing the LLM's parameters in the direction of duration consistency.

However, we cannot directly apply preference alignment algorithms (such as DPO \cite{17} or RLHF \cite{16-1,16-2}). This is primarily because the translation of each line of dialogue depends on its context, and the input to the SFT model needs to include multiple lines of dialogue. Consequently, fine-grained duration consistency alignment is required for each line of dialogue in the SFT model's response (see Table~\ref{tab:response}). Furthermore, further training of the SFT model must not alter the model's output format, as this could lead to issues such as translation omissions, resulting in synchronization problems with the timing notes of the original subtitles.

\begin{figure*}[t]
  \centering
  \includegraphics[width=\textwidth]{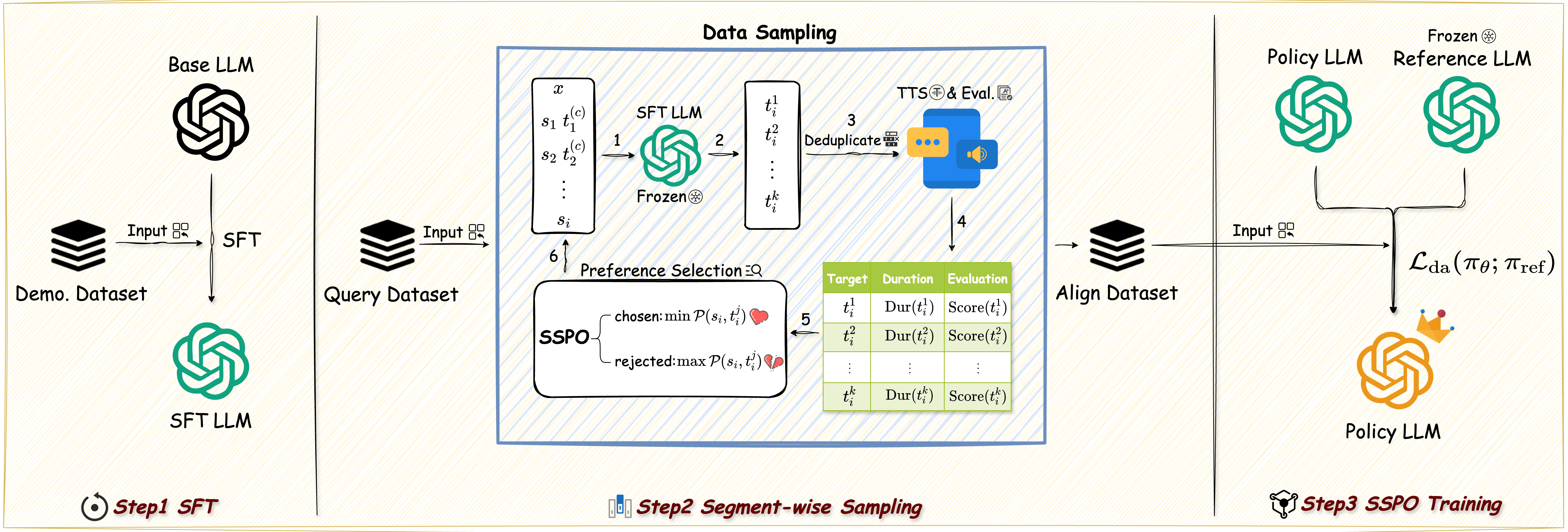}
  \caption{The overall framework of SSPO.}
  \label{fig:framework}
\end{figure*}

We present the overall framework of SSPO in Figure~\ref{fig:framework}. During DA, SSPO employs a fine-grained segment-wise sampling strategy to sample multiple translation candidates for each line of dialogue from SFT model. It then selects preferred and non-preferred translations for each line based on the duration consistency metric $\mathcal{P}(s_i,t_i)$. Subsequently, it optimizes the SFT model using a segment-wise DPO loss function. Furthermore, to ensure that the DA model does not deviate significantly from the SFT model and to maintain consistency in the model's output format, we incorporate a token-level KL divergence penalty term to constrain parameter updates during the training process.

\subsection{Sampling Strategy}

We first utilize the demonstration dataset to obtain a basic model $\pi _{\text{sft}}$ through SFT. For a sample $x\in \mathcal{D}_{\text{query}}$ from the query dataset $\mathcal{D}_{\text{query}}$ used for DA, which contains $n$ lines of dialogue $s_{1},s_{2},\dots ,s_{n}$ (see Table~\ref{tab:prompt} for examples). For each line $s_{i}$, we sample $k$ translation results based on the prefix $x,s_{1},t_{1}^{\text{(c)}},s_{2},t_{2}^{\text{(c)}},\dots ,s_{i-1},t_{i-1}^{\text{(c)}},s_{i}$, i.e., $\pi _{\text{sft}}(t_{i}|x,s_{1},t_{1}^{\text{(c)}},\dots ,s_{i-1},t_{i-1}^{\text{(c)}},s_{i})$, obtaining $\{t_{i}^{j}|j=1,2,\dots ,k\}$. After deduplication and discarding the bottom 20\% ranked by KIWI-XXL and XCOMET, we select the chosen translation $t_{i}^{\text{(c)}}$ and rejected translation $t_{i}^{\text{(r)}}$ based on the duration consistency metric $\mathcal{P}(s_i,t_i)$. Specifically, the sample with the minimum $\mathcal{P}(s_i,t_i)$ is chosen as $t_{i}^{\text{(c)}}$, and the one with the maximum as $t_{i}^{\text{(r)}}$ (see Table~\ref{tab:daexample} for demonstration). Finally, for the sample $x\in \mathcal{D}_{\text{query}}$, we obtain the corresponding sampling result $\mathcal{S}(x)\equiv \{(s_{i},t_{i}^{\text{(c)}},t_{i}^{\text{(r)}})|i=1,2,\dots ,n\}$. Algorithm~\ref{alg:sampling} provide a detailed illustration of the entire sampling process.

\begin{table*}[t]
\centering
\resizebox{\textwidth}{!}{
\begin{tabular}{lcccc}
 \Xhline{1.0pt}
 \rowcolor{gray!20}
 \textbf{Line} & \textbf{Duration(s)} & \textbf{Evaluation} & \textbf{Operation} \\
 \hline
 \begin{CJK}{UTF8}{gkai}历史虽然会重演，但是人类是无法回到过去的。\end{CJK} & 2.89 & - & - \\
 \hdashline
 \texttt{History repeats, but we can't go back to what was.}  & 2.66  & 85.6 & - \\
 \texttt{History might replay, but mankind cannot go back in time.} & 2.73  & 84.2 & discard \\
 \texttt{Even if history repeats, the past remains forever inaccessible to us.}  & \textbf{2.93}  & 89.3 & chosen \\
 \texttt{Although history may repeat itself, humans cannot return to the past.} & 3.03 & \textbf{91.4} & - \\
 \texttt{History often echoes, yet there’s no way for us to turn back the clock.} & 3.19 & 89.8 & rejected \\
 \Xhline{1.0pt}
\end{tabular}
}
\caption{Chosen and rejected translation selection. The evaluation score is the average of KIWI-XXL and XCOMET.}
\label{tab:daexample}
\end{table*}

\begin{algorithm}[!h]
  \caption{DPO Sampling Strategy.}
  \label{alg:sampling}
  \begin{algorithmic}[1]
    \Require SFT model $\pi _{\text{sft}}$, query dataset $\mathcal{D}_{\text{query}}$, sampling number $k$.
    \Ensure sampled sentence-level pairs set $\mathcal{S}(x)$.
    \State $//$ Iterate through the query dataset $\mathcal{D}_{\text{query}}$.
    \For {any $x\in \mathcal{D}_{\text{query}}$}
    \State $//$ Iterate through the dialogue lines in $x$.
    \For {$i=1$ to $n$}
    \State $//$ Sample multiple candidates.
    \For {$j=1$ to $k$}
    \State Sample $\pi _{\text{sft}}(t_{i}^{j}|\text{prefix})$
    \EndFor
    \State Deduplicate $\{t_{i}^{j}|j=1,\dots ,k\}$.
    \State Measure $\{t_{i}^{j}|j=1,\dots ,k\}$ by $\mathcal{P}$.
    \State Select chosen $t_{i}^{\text{(c)}}$ and rejected $t_{i}^{\text{(r)}}$.
    \EndFor
    \EndFor \\
    \Return $\mathcal{S}(x)\equiv \{(s_{i},t_{i}^{\text{(c)}},t_{i}^{\text{(r)}})|i=1,\dots ,n\}$.
  \end{algorithmic}
\end{algorithm}

The data sampling strategy in Algorithm~\ref{alg:sampling} is predicated on the generation diversity of dialogue translations. Specifically, LLMs typically generate various translations $t_{i}$ for most lines $s_{i}$, with each $t_{i}$ having a distinct duration. However, simple lines such as "Good morning" and "How are you?" with lower translation diversity should not be utilized for model optimization. Furthermore, if the duration differences among various translations $t_{i}$ are insignificant, they contribute little to model optimization. Consequently, based on the sampled data, we establish two thresholds, $\varepsilon _{1}$ and $\varepsilon _{2}$, to filter out lines with low translation diversity. Specifically, if the number of deduplicated samples from $k$ samples of $s_{i}$ is less than $\varepsilon _{1}$ or $\mathcal{P}(s_{i},t_{i}^{\text{(r)}})-\mathcal{P}(s_{i},t_{i}^{\text{(c)}})< \varepsilon _{2}$, the line $s_{i}$ will not be involved in the preference optimization process. In our experiment, we set $\varepsilon _{1}=4$ and $\varepsilon _{2}=0.08$. Ultimately, we obtain the dataset $\mathcal{D}_{\text{dpo}}$ used for DPO training.

\subsection{Alignment Loss Optimization}

Unlike preference alignment tasks for language models, DA task requires fine-grained alignment of multiple segments within the LLM's response, rather than aligning the entire response as in preference alignment. Additionally, due to the contextual dependencies in dialogue translation, DA must ensure that the LLM output format (see Table~\ref{tab:response}) remains unchanged to prevent interference with the correspondence between the original line and its translation. SSPO utilizes DPO loss and sampled data to achieve fine-grained alignment of the duration for each line of dialogue. SSPO similarly requires the scheduling of two models: the policy $\pi _{\theta }$ and the reference $\pi _{\text{ref}}$, both of which are initialized from the SFT model $\pi _{\text{sft}}$. Specifically, for a sample $(x,\mathcal{S}(x))\in \mathcal{D}_{\text{dpo}}$ from the sampled DPO dataset $\mathcal{D}_{\text{dpo}}$, we employ the standard DPO loss \cite{17} to calculate a DPO loss term for a single line of dialogue $s_{i}$ based on $(s_{i},t_{i}^{\text{(c)}},t_{i}^{\text{(r)}})$:
\begin{equation}
\resizebox{0.43\textwidth}{!}{$
\begin{aligned}
&\mathcal{L}_{\text{dpo}}(s_{i})=\\&\text{log}\, \sigma \left (\beta \, \text{log}\frac{\pi _{\theta }(t_{i}^{\text{(c)}}|p_{i})}{\pi _{\text{ref}}(t_{i}^{\text{(c)}}|p_{i})}-\beta \, \text{log}\frac{\pi _{\theta }(t_{i}^{\text{(r)}}|p_{i})}{\pi _{\text{ref}}(t_{i}^{\text{(r)}}|p_{i})}\right ),
\end{aligned}
$}
\end{equation}
where $p_{i}$ is $x,s_{1},t_{1}^{\text{(c)}},s_{2},t_{2}^{\text{(c)}},\dots ,s_{i-1},t_{i-1}^{\text{(c)}},s_{i}$, and $\beta$ is a hyperparameter used to control the sensitivity of the optimization process to reward differences. $\mathcal{L}_{\text{dpo}}(s_{i})$ only controls the duration of $t_{i}$ without affecting other lines in $x$, thereby achieving independent customized DA for each line. We can now derive the loss function for DA as follows:
\begin{equation}
\begin{split}
&\mathcal{L}_{\text{da}}(\pi _{\theta };\pi _{\text{ref}})=\\&-\mathbb{E}_{(x,\mathcal{S}(x))\sim \mathcal{D}_{\text{dpo}}}\left (\sum_{i=1}^{n}\mathcal{L}_{\text{dpo}}(s_{i})\right ).
\end{split}
\end{equation}

\subsection{Output Format Control}
\label{sec:ofc}

DA, as a CTG task, utilizes $\mathcal{L}_{\text{da}}(\pi _{\theta };\pi _{\text{ref}})$ to achieve precise control over the duration of translated dialogue. However, it fails to maintain the consistent output format of LLMs. This limitation primarily stems from the fact that vanilla DPO is designed for open-ended generation tasks \cite{17,sdpo}, relying on sentence-level KL divergence constraints, which are negligible for generation tasks with fixed output formats. Consequently, DA requires more stringent constraints. We adapt two methods to constrain the generation format: Token-level KL Divergence (TKLD) constraints and Low-Rank Adaptation (LoRA) training \cite{lora}.

\subsubsection{Token-level KL Divergence}

TKLD constraint is employed to regulate the token generation distribution output by the policy model $\pi _{\theta }$. During the model optimization process, this constraint ensures that the output distribution of $\pi _{\theta }$ remains as consistent as possible with that of the reference model $\pi _{\text{ref}}$. This approach not only guarantees the consistency of output formats between $\pi _{\theta }$ and $\pi _{\text{ref}}$, but also prevents $\pi _{\theta }$ from deviating too far from $\pi _{\text{ref}}$, thereby ensuring that the translation quality of the model after DA does not significantly deteriorate. The loss function incorporating the TKLD constraint is as follows:
\begin{equation}
\begin{split}
&\mathcal{L}_{\text{tkld}}(\pi _{\theta };\pi _{\text{ref}})=\mathcal{L}_{\text{da}}(\pi _{\theta };\pi _{\text{ref}})\\&+\lambda \cdot \sum _{t}\text{KL}(\pi _{\theta }(\cdot |x,y_{t}),\pi _{\text{ref}}(\cdot |x,y_{t})),
\end{split}
\end{equation}
where $\lambda$ is a hyperparameter used to control the constraint strength, and $y_t$ represents all tokens generated at step $t$.

\subsubsection{Low-Rank Adaptation Training}

In addition to TKLD constraint, employing LoRA training can also help maintain the output format of the policy model, preventing model collapse while significantly reducing computational resource requirements during the training process. However, the LoRA training process converges more slowly compared to full-parameter training, necessitating a greater number of training iterations.

\section{Experiments}

\subsection{Experimental Settings}

We use our custom PolySC dataset for \texttt{zh}$\Rightarrow$\texttt{en} and \texttt{zh}$\Rightarrow$\texttt{th} translation experiments. Each direction's dataset is further divided into Demonstration and Query datasets for SFT and DA training. Additionally, we reserve 4 television series for the test set, ensuring these data are not present in the training set. Details of PolySC is shown in Appendix~\ref{sec:dataset}.

We compare SSPO with the following baselines:
\begin{itemize}
\item \textbf{AutoDubbing} \cite{2} models isochrony by controlling verbosity of NMT.
\item \textbf{VideoDubber} \cite{1} constructs a speech-aware length-controlled NMT model.
\item \textbf{GPT-3.5-Turbo} is an early chat language model released by OpenAI (\texttt{gpt-3.5-turbo} \texttt{-0125}).
\item \textbf{GPT-4o}\footnote{\url{https://platform.openai.com/docs/models}} is OpenAI's current most advanced multimodal model (\texttt{gpt-4o-2024-11-20}).
\item \textbf{Claude 3.5 Sonnet}\footnote{\url{https://docs.anthropic.com/en/api}} is a multimodal model released by Anthropic in June 2024.
\item \textbf{Llama3.1-8B-Chinese-Chat}\footnote{\url{https://huggingface.co/shenzhi-wang/Llama3.1-8B-Chinese-Chat}}, \textbf{GLM-4-9B-Chat} \cite{glm}, and \textbf{Qwen2.5-14B-Instruct} \cite{qwen25} are open-source language models released by Meta Platforms Inc., Zhipu AI, and Alibaba Group respectively, used as foundation models for SSPO.
\end{itemize}

For detailed experimental settings, refer to Appendix~\ref{sec:expdetail}. The source code for SSPO's data sampling and training is available at \url{https://github.com/CcQunResearch/SSPO}.

\subsection{Results and Discussion}

In this subsection, we present SSPO evaluation results, visualizations, and case studies.

\subsubsection{Main Experiments}

In Table~\ref{tab:exp-main}, we present the main evaluation experiments for SSPO, reporting six metrics: S>T Rate, S>T Dur, T>S Rate, T>S Dur, Consistency Rate (CR), and $\mathcal{P}$. These metrics respectively represent the proportion of lines where the source duration exceeds the target duration by more than 0.1s and the average excess duration (s), the proportion of lines where the target duration exceeds the source duration by more than 0.1s and the average excess duration (s), the proportion of lines where the difference between source and target durations is within 0.1s, and the average value of the duration consistency metric $\mathcal{P}$. Additionally, we compare our results with the Gold Reference human translations from the test set and the Alignment Bound of DA. It is important to note that while DA aims to minimize $\mathcal{P}$ by making the duration of translated lines as consistent as possible with the source lines, achieving perfect consistency ($\mathcal{P}=0$) is infeasible. This is because, when maintaining translation quality, the most duration-consistent translation for each line typically does not yield a $\mathcal{P}$ of 0 with the source (see the example in Table~\ref{tab:daexample}). Consequently, DA has an upper limit, termed the Alignment Bound, which is inaccessible. However, we can estimate this bound by calculating the average $\mathcal{P}$ between all chosen translations and their corresponding source lines in the data sampled using Algorithm~\ref{alg:sampling}. 

\begin{table*}[!h]
\centering
\resizebox{\textwidth}{!}{
\begin{tabular}{cccccccc|cccccc}
 \Xhline{1.0pt}
 \rowcolor{gray!20}
 ~ & ~ & \multicolumn{6}{c}{\textbf{\texttt{zh}$\Rightarrow$\texttt{en}}} & \multicolumn{6}{c}{\textbf{\texttt{zh}$\Rightarrow$\texttt{th}}}\\
 \cline{3-14}
 \rowcolor{gray!20}
 \multirow{-2}{*}{\textbf{Method}} & \multirow{-2}{*}{\textbf{Train}} & \textbf{S>T Rate} & \textbf{S>T Dur} & \textbf{T>S Rate} & \textbf{T>S Dur} & \textbf{CR} & $\mathcal{P}$ & \textbf{S>T Rate} & \textbf{S>T Dur} & \textbf{T>S Rate} & \textbf{T>S Dur} & \textbf{CR} & $\mathcal{P}$\\
 \hline
 Gold Reference & - & \textit{18.0\%} & \textit{0.344} & \textit{64.1\%} & \textit{0.464} & \textit{17.9\%} & \textit{0.501} & \textit{19.4\%} & \textit{0.369} & \textit{60.2\%} & \textit{0.460} & \textit{20.3\%} & \textit{0.489} \\
 \hdashline
 AutoDubbing & SFT & 22.6\% & 0.355 & 56.4\% & 0.400 & 21.0\% & 0.388 & 20.5\% & 0.300 & 51.7\% & 0.388 & 27.8\% & 0.334 \\
 VideoDubber & SFT & 26.5\% & 0.363 & 51.3\% & 0.371 & 22.2\% & 0.344 & 23.3\% & 0.305 & 47.8\% & 0.369 & 29.0\% & 0.314 \\ 
 \hdashline
 GPT-3.5-Turbo & PE & \colorbox{mlb}{\textbf{9.2\%}} & \colorbox{mlb}{\textbf{0.267}} & 73.1\% & 0.465 & 17.7\% & 0.526 & \colorbox{mlb}{\textbf{14.2\%}} & 0.302 & 65.2\% & 0.487 & 20.7\% & 0.567 \\
 GPT-4o & PE & 14.7\% & 0.295 & 66.1\% & 0.407 & 19.2\% & 0.417 & 18.2\% & 0.299 & 54.9\% & 0.353 & 27.0\% & 0.318 \\
 Claude 3.5 Sonnet & PE & \colorbox{mlo}{\textbf{11.4\%}} & \colorbox{mlb}{\textbf{0.267}} & 69.3\% & 0.401 & 19.3\% & 0.410 & \colorbox{mlo}{\textbf{14.8\%}} & \colorbox{mlb}{\textbf{0.274}} & 57.7\% & 0.342 & 27.5\% & 0.313 \\
 \hdashline
 \multirow{2}{*}{Llama3.1-8B-CN-Chat} & SFT & 22.7\% & 0.352 & 56.6\% & 0.398 & 20.7\% & 0.389 & 17.2\% & 0.287 & 56.4\% & 0.410 & 26.4\% & 0.370 \\
 ~ & \textbf{SSPO} & 31.7\% & 0.341 & \colorbox{mlo}{\textbf{42.4\%}} & \colorbox{mlb}{\textbf{0.309}} & \colorbox{mlb}{\textbf{25.9\%}} & \colorbox{mlb}{\textbf{0.263}} & 30.7\% & \colorbox{mlo}{\textbf{0.285}} & \colorbox{mlo}{\textbf{32.9\%}} & \colorbox{mlb}{\textbf{0.279}} & \colorbox{mlb}{\textbf{36.4\%}} & \colorbox{mlo}{\textbf{0.206}} \\
 \hdashline
 \multirow{2}{*}{GLM-4-9B-Chat} & SFT & 19.5\% & 0.342 & 60.5\% & 0.427 & 20.0\% & 0.428 & 18.1\% & 0.291 & 55.0\% & 0.391 & 27.0\% & 0.360 \\
 ~ & \textbf{SSPO} & 29.6\% & 0.350 & 45.9\% & \colorbox{mlo}{\textbf{0.323}} & 24.5\% & 0.283 & 25.9\% & 0.291 & 42.0\% & 0.318 & 32.1\% & 0.254 \\
 \hdashline
 \multirow{2}{*}{Qwen2.5-14B-Instruct} & SFT & 20.0\% & 0.341 & 59.8\% & 0.439 & 20.2\% & 0.423 & 18.2\% & 0.294 & 55.4\% & 0.397 & 26.4\% & 0.362 \\
 ~ & \textbf{SSPO} & 34.4\% & 0.366 & \colorbox{mlb}{\textbf{40.6\%}} & 0.324 & \colorbox{mlo}{\textbf{24.9\%}} & \colorbox{mlo}{\textbf{0.272}} & 38.6\% & 0.290 & \colorbox{mlb}{\textbf{25.3\%}} & \colorbox{mlb}{\textbf{0.279}} & \colorbox{mlo}{\textbf{36.1\%}} & \colorbox{mlb}{\textbf{0.198}} \\
 \hdashline
 Alignment Bound & - & \textit{16.4\%} & \textit{0.278} & \textit{39.3\%} & \textit{0.331} & \textit{44.3\%} & \textit{0.220} & \textit{9.2\%} & \textit{0.232} & \textit{40.4\%} & \textit{0.313} & \textit{50.4\%} & \textit{0.203} \\
 \Xhline{1.0pt}
\end{tabular}}
\caption{\texttt{zh}$\Rightarrow$\texttt{en} and \texttt{zh}$\Rightarrow$\texttt{th} results on test set. The best and second best results are denoted as \colorbox{mlb}{\textbf{blue}} and \colorbox{mlo}{\textbf{orange}}.}
\label{tab:exp-main}
\end{table*}

Results in Table~\ref{tab:exp-main} demonstrate that after SSPO training, SFT models show a significant decrease in $\mathcal{P}$ and a notable increase in the dialogue duration consistency rate, outperforming other baselines. SSPO produces consistent alignment effects across different base models, validating its universal applicability. GPT-3.5, GPT-4, and Claude 3.5, which use PE to control translation duration (see Appendix~\ref{sec:pe} for the prompt design), showed some improvement compared to the gold reference, but failed to match the performance of traditional methods like AutoDubbing and VideoDubber. This indicates that LLMs inherently lack sufficient perception of text duration and require additional duration information to effectively complete DA tasks. For LLMs, it is challenging to design a differentiable loss function that incorporates extra duration information to directly optimize them without modifying their underlying embeddings, model architecture, or introducing additional model parameters. Therefore, SSPO approaches this as a preference optimization problem, achieving DA through fine-grained sampling and training. 

In addition, we conduct experiments on Spanish-related \texttt{zh}$\Rightarrow$\texttt{es} and \texttt{es}$\Rightarrow$\texttt{zh} translations in Appendix~\ref{sec:expes}, which also demonstrate similar performance. And we explore two other alternative solutions for DA in Appendix~\ref{sec:furtherexp}. 

\subsubsection{Visualization and Case Study}

\begin{figure*}[!h]
  \centering
  \subfigure[\texttt{zh}$\Rightarrow$\texttt{en}]{\includegraphics[width=0.48\textwidth]{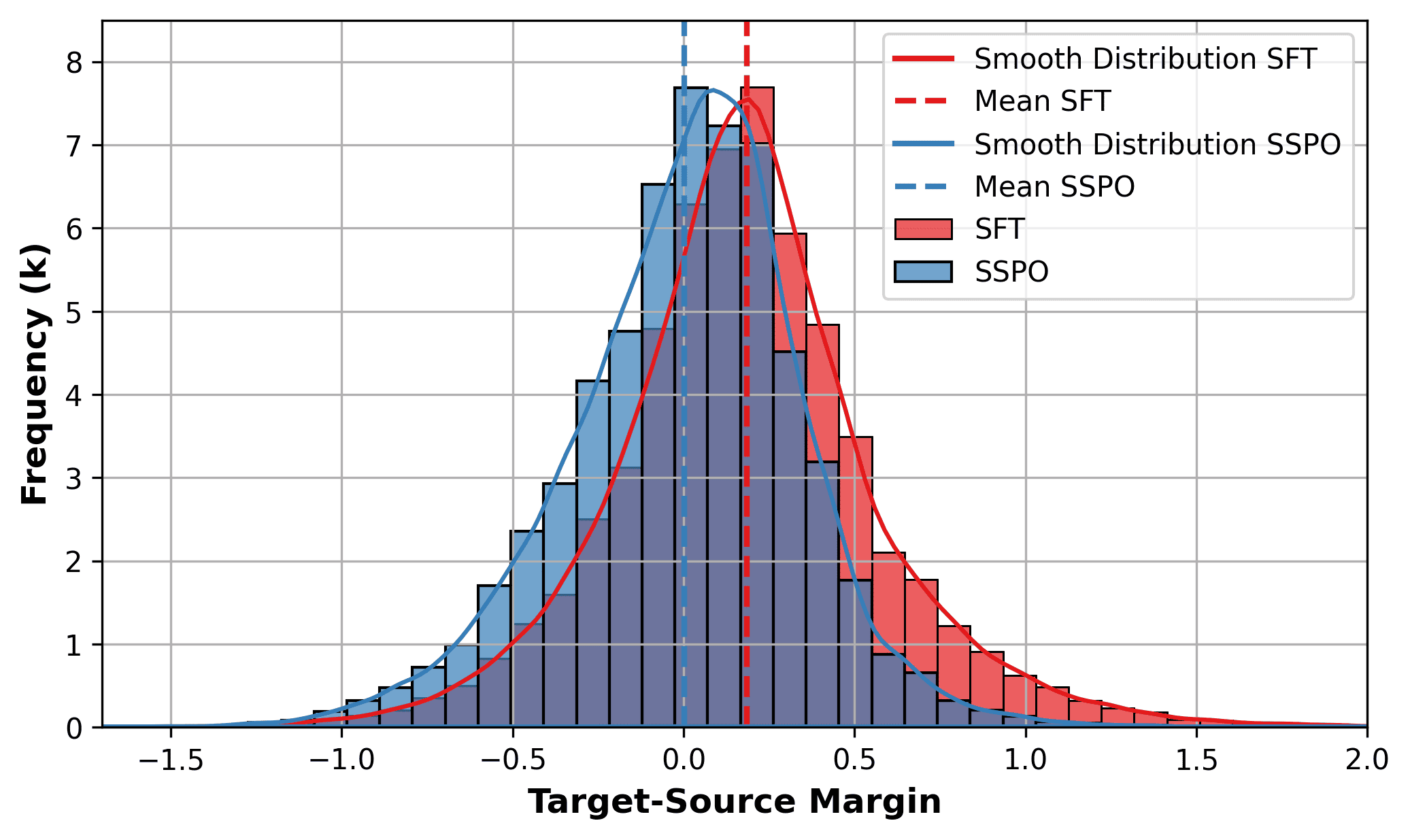}}
  \subfigure[\texttt{zh}$\Rightarrow$\texttt{th}]{\includegraphics[width=0.48\textwidth]{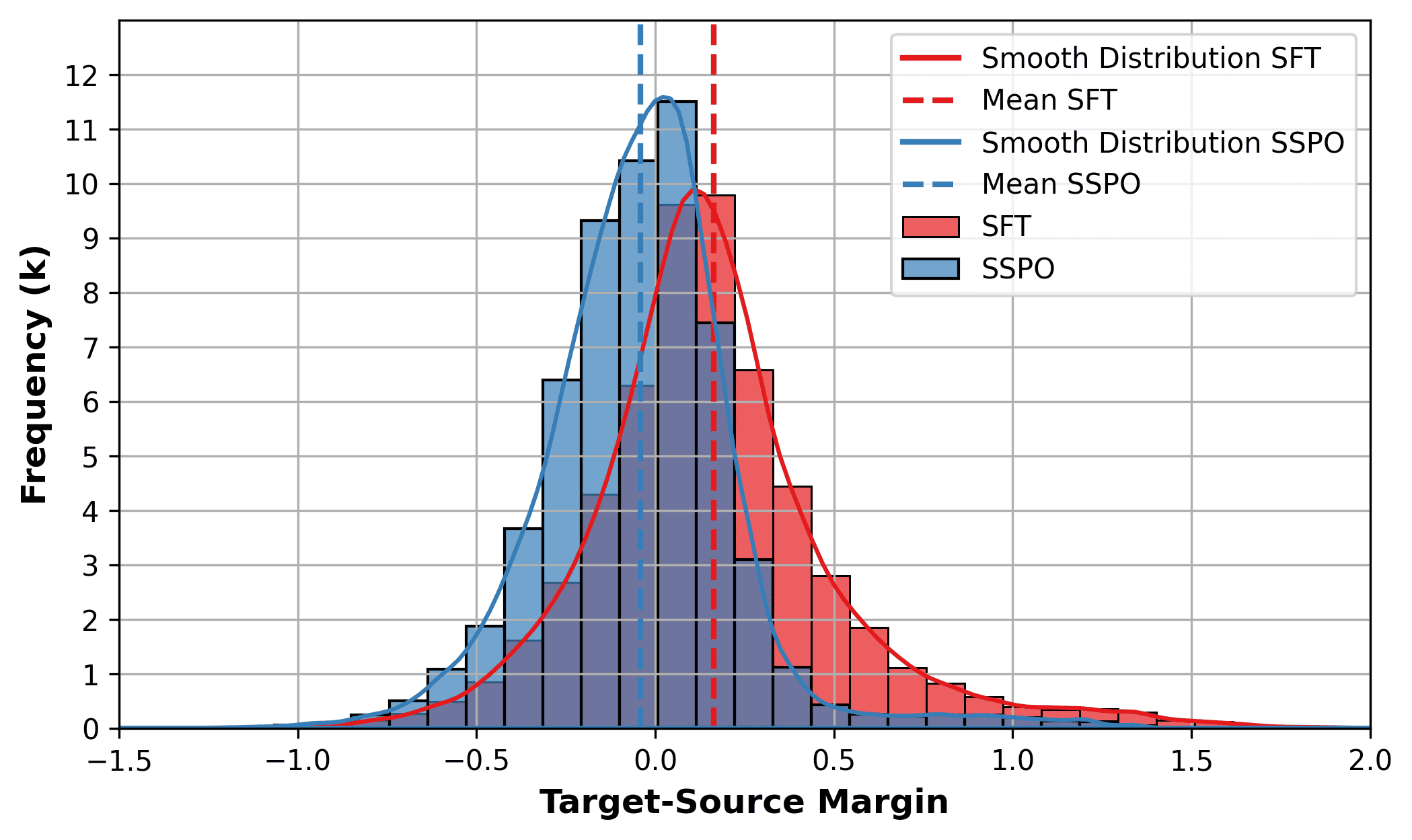}}
  \caption{Frequency distribution of Qwen2.5-14B-Instruct model on \texttt{zh}$\Rightarrow$\texttt{en} and \texttt{zh}$\Rightarrow$\texttt{th} translations.}
  \label{fig:visenth}
\end{figure*}

In Figure~\ref{fig:visenth}, we present the frequency distribution of the duration difference between the translation and the original text for both SFT and SSPO models of Qwen2.5-14B-Instruct, to observe the changes in translation duration after SSPO alignment. It is evident that after SSPO training, the duration discrepancy between the original text and the translation significantly narrows. This is reflected in Figure~\ref{fig:visenth}, where the histogram for SSPO is noticeably more concentrated around zero compared to that of SFT. Additionally, in Table~\ref{tab:casestudy}, we showcase comparative case studies of translations for certain lines by the SFT and SSPO models of Qwen2.5-14B-Instruct. These examples visually demonstrate that the translations aligned by SSPO exhibit greater duration consistency with the original text compared to those generated by the SFT model.

\begin{table*}[!h]
\centering
\resizebox{0.99\textwidth}{!}{
\begin{tabular}{llll}
 \Xhline{1.0pt}
 \rowcolor{gray!20}
 \textbf{Type} & \textbf{Source} & \textbf{Target} & \textbf{Model}\\
 \hline
 \multirow{2}{*}{T>S} & \multirow{2}{*}{\begin{CJK}{UTF8}{gkai}灵王的交代您还记得吧？\end{CJK} (1.52s)} & \texttt{Do you still remember what the Spirit King told you?} (2.37s) & SFT\\
 ~ & ~ & \texttt{Do you remember what the Spirit King said?} (1.97s) & SSPO\\
 \hdashline
 \multirow{2}{*}{T>S} & \multirow{2}{*}{\begin{CJK}{UTF8}{gkai}你们摆平那群玄门人了？\end{CJK} (1.53s)} & \texttt{Have you dealt with those arcanists?} (1.61s) & SFT\\
 ~ & ~ & \texttt{You've dealt with those arcanists?} (1.55s) & SSPO\\
 \hdashline
 \multirow{2}{*}{T>S} & \multirow{2}{*}{\begin{CJK}{UTF8}{gkai}或许你我可以尝试做一对有情人\end{CJK} (1.75s)} & \texttt{Maybe you and I can try to be a pair of lovers.} (2.32s) & SFT\\
 ~ & ~ & \texttt{Maybe you and I can be that lovebird.} (1.86s) & SSPO\\
 \hdashline
 \multirow{2}{*}{S>T} & \multirow{2}{*}{\begin{CJK}{UTF8}{gkai}必须赶紧去取幽冥赋了\end{CJK} (1.82s)} & \texttt{I must go and get the Ghost Charmer.} (1.7s) & SFT\\
 ~ & ~ & \texttt{I have to retrieve the Ghost Charmer soon.} (1.83s) & SSPO\\
 \hdashline
 \multirow{2}{*}{S>T} & \multirow{2}{*}{\begin{CJK}{UTF8}{gkai}这中餐合你胃口吗？\end{CJK} (1.44s)} & \texttt{Do you like Chinese food?} (1.12s) & SFT\\
 ~ & ~ & \texttt{Is Chinese cuisine palatable to you?} (1.54s) & SSPO\\
 \Xhline{1.0pt}
\end{tabular}
}
\caption{Case studies of Qwen2.5-14B-Instruct model on \texttt{zh}$\Rightarrow$\texttt{en} translation.}
\label{tab:casestudy}
\end{table*}

\subsection{Human Evaluation of Translation Quality}
\label{sec:humaneval}

We did not utilize traditional translation quality evaluation metrics such as BLEU and ROUGE. These metrics overlook the semantics of the translation, lack contextual understanding, and cannot handle the diversity and flexibility of LLM translations. Therefore, we completely abandoned these metrics in favor of human evaluation. In Table~\ref{tab:humaneval}, we present the human evaluation results for the translation quality of SSPO. We conducted evaluations in both the \texttt{zh}$\Rightarrow$\texttt{en} and \texttt{es}$\Rightarrow$\texttt{zh} directions. For each direction, we employed four evaluators, all of whom are bachelor's or master's degree professionals in English or Spanish translation, with Chinese as their native language. Due to the subjective preferences of different evaluators, we did not use scoring in the human evaluation. Instead, we performed pairwise comparisons of different translations to assess the win rate metric. 

\begin{table*}[!h]
\centering
\resizebox{\textwidth}{!}{
\begin{tabular}{clcccc|cccc}
\Xhline{1.0pt}
\rowcolor{gray!20}
~ & ~ & \multicolumn{4}{c}{\texttt{zh}$\Rightarrow$\texttt{en}} & \multicolumn{4}{c}{\texttt{es}$\Rightarrow$\texttt{zh}} \\
\cline{3-10}
\rowcolor{gray!20}
\multirow{-2}{*}{\textbf{Challenger}} & \multirow{-2}{*}{\textbf{Competitors}} & \textbf{Accuracy} & \textbf{Naturalness} & \textbf{Vividness} & \textbf{Comprehensive} & \textbf{Accuracy} & \textbf{Naturalness} & \textbf{Vividness} & \textbf{Comprehensive} \\
\hline
\multirow{4}{*}{\makecell{SSPO\\(GLM-4-9B)}} & Gold Reference & \colorbox{mlb}{27:50:23} & \colorbox{mlo}{16:67:17} & \colorbox{mlo}{23:51:26} & \colorbox{mlo}{24:45:31} & \colorbox{mlo}{19:61:20} & \colorbox{mlb}{23:55:22} & \colorbox{mlo}{24:45:31} & \colorbox{mlo}{24:48:28} \\
~ & Vanilla GLM-4-9B  & \colorbox{mlb}{26:50:24} & \colorbox{mlb}{21:60:19} & \colorbox{mlb}{24:55:21} & \colorbox{mlb}{27:50:23} & \colorbox{mlb}{28:51:21} & \colorbox{mlb}{26:52:22} & \colorbox{mlb}{27:51:22} & \colorbox{mlb}{28:53:19}\\
~ & GPT-4o & \colorbox{mlo}{23:50:27} & \colorbox{mlo}{19:57:24} & \colorbox{mlo}{23:51:26} & \colorbox{mlo}{20:52:28} & \colorbox{mlb}{23:55:22} & \colorbox{mlo}{19:58:23} & \colorbox{mlo}{20:56:24} & \colorbox{mlo}{22:52:26}\\
~ & SFT & \colorbox{mlo}{23:52:25} & \colorbox{mlo}{18:60:22} & \colorbox{mlo}{24:49:27} & \colorbox{mlo}{29:40:31} & \colorbox{mlb}{23:57:20} & \colorbox{mlb}{23:56:22} & \colorbox{mlo}{27:45:28} & \colorbox{mlb}{27:51:22} \\
\hline
\multirow{4}{*}{\makecell{SSPO\\(Qwen2.5-14B)}} & Gold Reference & \colorbox{mlb}{21:63:16} & \colorbox{mlb}{24:55:21} & \colorbox{mlo}{23:51:26} & \colorbox{mlo}{24:50:26} & \colorbox{mlb}{21:59:20} & \colorbox{mlb}{24:55:21} & \colorbox{mlo}{25:48:27} & \colorbox{mlb}{27:49:24} \\
~ & Vanilla Qwen2.5-14B & \colorbox{mlo}{24:51:25} & \colorbox{mlb}{26:50:24} & \colorbox{mlb}{26:53:21} & \colorbox{mlb}{32:41:27} & \colorbox{mlb}{31:45:24} & \colorbox{mlb}{26:51:23} & \colorbox{mlb}{27:48:25} & \colorbox{mlb}{30:43:27} \\
~ & GPT-4o & \colorbox{mlb}{25:51:24} & \colorbox{mlo}{19:58:22} & \colorbox{mlo}{23:48:29} & \colorbox{mlo}{23:49:28} & \colorbox{mlb}{25:53:22} & \colorbox{mlb}{27:51:22} & \colorbox{mlo}{24:50:26} & \colorbox{mlb}{31:43:26}\\
~ & SFT & \colorbox{mlo}{23:48:29} & \colorbox{mlo}{21:54:25} & \colorbox{mlo}{24:49:27} & \colorbox{mlo}{24:50:26} & \colorbox{mlb}{27:49:24} & \colorbox{mlb}{33:38:28} & \colorbox{mlb}{25:51:24} & \colorbox{mlb}{30:46:24} \\
\Xhline{1.0pt}
\end{tabular}
}
\caption{Human translation quality evaluation, reporting win rate (win:tie:loss). The winning and losing contrasts are marked in \colorbox{mlb}{blue} and \colorbox{mlo}{orange}, respectively.}
\label{tab:humaneval}
\end{table*}

The SSPO model was evaluated against four baselines: the gold reference, vanilla base model, GPT-4o, and the SFT model, across the dimensions of accuracy, naturalness, and vividness: 1) \textit{Accuracy}: Does the translation accurately convey the original meaning of the dialogue? 2) \textit{Naturalness}: Is the translation fluent and does it conform to the grammar and lexical conventions of the target language? 3) \textit{Vividness}: Is the translation expressive and does it convey the emotion and ambiance of the original dialogue? Additionally, we conducted a comprehensive evaluation, with instructions provided to evaluators as referenced in Appendix~\ref{sec:ioqe}, similar to those for other dimensions. The multidimensional evaluation across both directions comprised a total of 64 evaluation tasks, with each evaluator randomly assigned 8 tasks. In each evaluation task, we provided evaluators with challenger and competitor translations of 200 dialogue segments from the test set, each segment containing 20 lines of dialogue. Evaluators were required to select the superior translation or mark both as "no significant difference." The dialogue segments for different evaluation tasks were randomly selected subsets from the test set.

Translations by LLM-based methods, like SSPO and GPT-4o, often surpass human in terms of accuracy but fall short in vividness. Human translators typically reference scene and emotional cues from the video and audio, which LLMs are currently unable to incorporate. Future research should focus on enhancing translation models' ability to perceive and understand multimodal information to achieve more vivid localized translations. SSPO shows significant improvement over its vanilla base model, demonstrating the positive impact of fine-tuning LLMs on visual media data. The comparison between SSPO and SFT model further highlights the influence of SSPO on model performance. In translations from high-to-low information density, SSPO usually reduces the length of generated translations, which inevitably results in some information loss as the model sacrifices some translation quality for better duration control. Conversely, in low-to-high information density translations, the model tends to generate longer (more informative) translations, thereby improving translation quality. Unlike the strict accuracy requirements in legal text translations, subtitle translation can tolerate some loss of accuracy because viewers can rely on other modalities, such as video and audio, to supplement their understanding of the program's current scene, even if some information is lost in the translation.

\subsection{Ablation Study}

We conducted a series of ablation studies to investigate the impact of various factors on SSPO.

\subsubsection{Format Control Measures}

As a CTG task, DA requires precise control over the duration of each line while also ensuring that the model's output adheres to the format shown in Table~\ref{tab:response}. In Section~\ref{sec:ofc}, we proposed two format control methods: TKLD and LoRA training, and compared the three configurations of full parameter fine-tuning, TKLD, and LoRA training in Table~\ref{tab:format} in terms of the output conforming to the required format. We conducted experiments with two models on translation tasks in two languages, where the efficient rate represents the proportion of lines in the test set that conform to the format. The results show that full parameter fine-tuning leads to a significant drop in the efficient rate, and the model may encounter issues such as complete output collapse, omission of certain lines, or failure to adhere to the required format. Both LoRA and TKLD are able to maintain the output format after SSPO alignment, with LoRA achieving an efficient rate close to 100\%. Moreover, LoRA requires less GPU memory compared to TKLD. Therefore, we recommend using LoRA for SSPO training.

\begin{table}[!h]
\centering
\resizebox{0.49\textwidth}{!}{
\begin{tabular}{cccc|cc}
 \Xhline{1.0pt}
 \rowcolor{gray!20}
 ~ & ~ & \multicolumn{2}{c}{\textbf{\texttt{zh}$\Rightarrow$\texttt{en}}} & \multicolumn{2}{c}{\textbf{\texttt{zh}$\Rightarrow$\texttt{th}}}\\
 \cline{3-6}
 \rowcolor{gray!20}
 \multirow{-2}{*}{\textbf{Base Model}} & \multirow{-2}{*}{\textbf{Train}} & \textbf{Efficient Rate} & $\mathcal{P}$ & \textbf{Efficient Rate} & $\mathcal{P}$ \\
 \hline
 \multirow{3}{*}{Llama3.1-8B} & - & 89.6\% & \textbf{0.251} & 85.1\% & \textbf{0.197} \\
 ~ & TKLD & 98.1\% & 0.273 & 97.4\% & 0.203 \\
 ~ & LoRA & \textbf{99.7\%} & 0.263 & \textbf{99.8\%}  & 0.206 \\
 \hdashline
 \multirow{3}{*}{Qwen2.5-14B} & - & 81.2\% & \textbf{0.258} & 73.9\% & 0.202 \\
 ~ & TKLD & 96.9\% & 0.283 & 97.2\% & 0.209 \\
 ~ & LoRA & \textbf{99.8}\% & 0.272 & \textbf{99.7\%}  & \textbf{0.198} \\
 \Xhline{1.0pt}
\end{tabular}
}
\caption{The impact of format control measures.}
\label{tab:format}
\end{table}

\subsubsection{Reward Difference}

\begin{figure}[!h]
  \centering
  \includegraphics[width=0.48\textwidth]{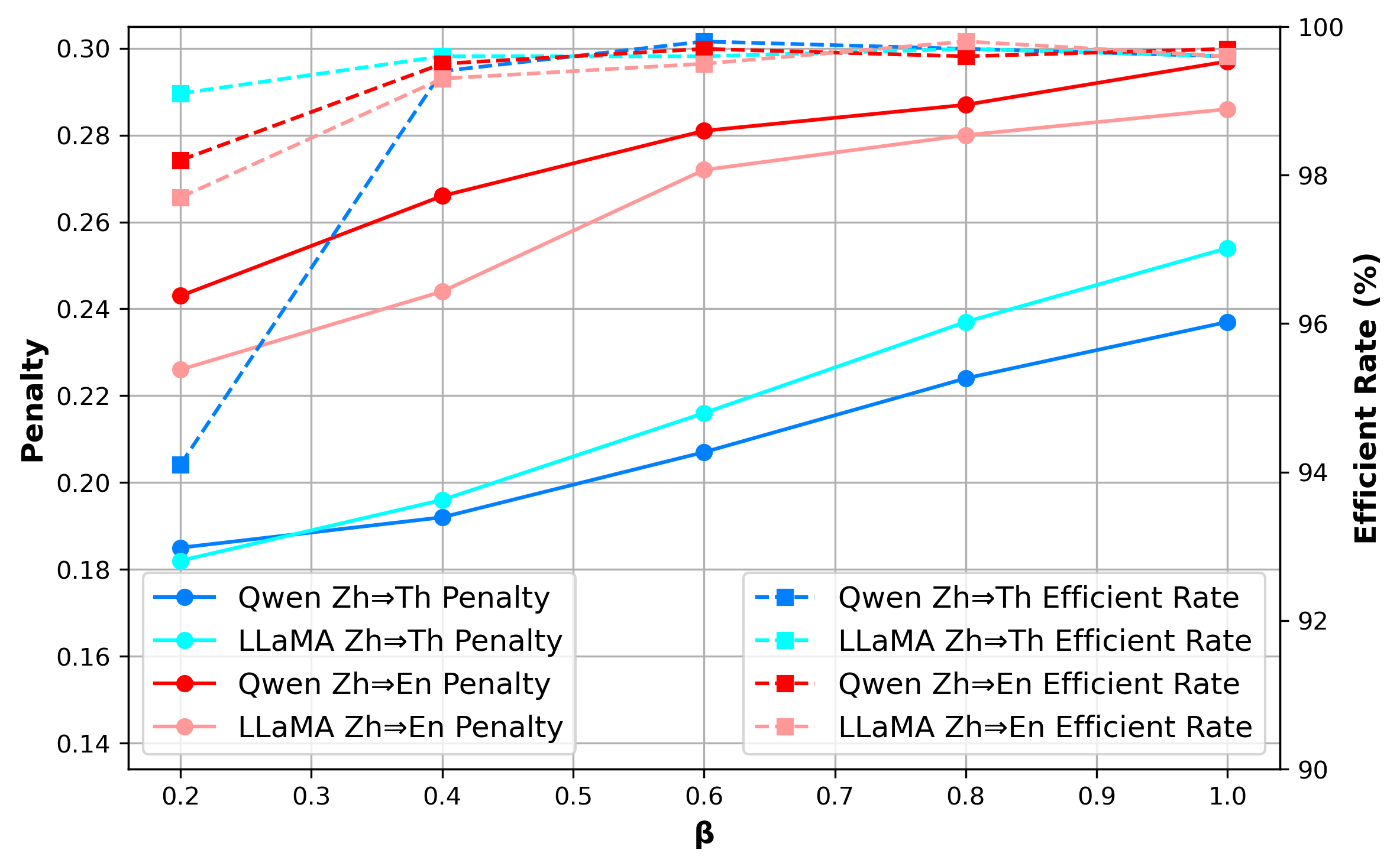}
  \caption{The impact of hyperparameter $\beta$.}
  \label{fig:beta}
\end{figure}

The hyperparameter $\beta$ in SSPO loss controls the model's sensitivity to implicit reward differences. We investigated the impact of $\beta$ on DA performance using two models for translation in two languages. Figure~\ref{fig:beta} reports the changes in the duration consistency metric $\mathcal{P}$ and the format efficiency rate as $\beta$ varies. The results indicate that as $\beta$ increases, $\mathcal{P}$ steadily increases, suggesting that smaller values of $\beta$ lead to higher duration consistency. However, when using smaller $\beta$ values, occasional decreases in the format efficiency rate were observed, with the model showing instances of non-adherence to the output format. Therefore, considering these factors comprehensively, we opted for a moderate value of $\beta=0.5$ in our experiments.

\subsubsection{Data Scale}

\begin{figure}[!h]
  \centering
  \includegraphics[width=0.48\textwidth]{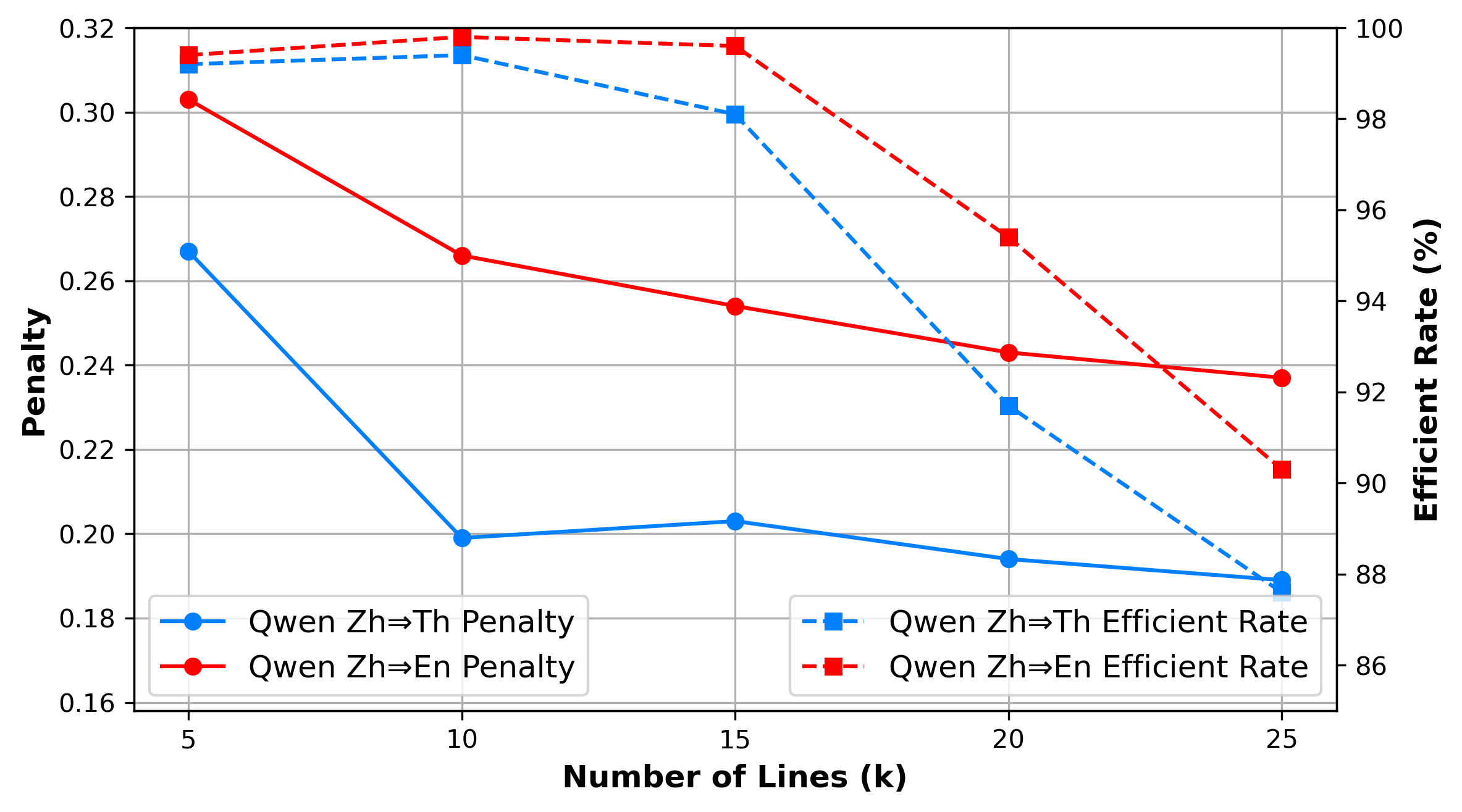}
  \caption{The impact of data scale.}
  \label{fig:datascale}
\end{figure}

Another question worth exploring is "How much data does SSPO require to achieve acceptable DA performance?" To address this, we investigated the impact of the number of dialogue lines from the Query dataset on performance using Qwen2.5-14B-Instruct for translation in two languages. Figure~\ref{fig:datascale} reports the changes in the duration consistency metric $\mathcal{P}$ and the format efficiency rate as the data scale varies. The results show that as the data scale increases, $\mathcal{P}$ gradually decreases, indicating that using more data yields better alignment effects. However, employing an excessive amount of data leads to a sharp decline in the format efficiency rate. Considering these factors, we used approximately 10,000 dialogue lines in our experiments, which is equivalent to about 600 prompt-response pairs from the Query dataset, representing roughly 3\% of the entire PolySC dataset. This approach achieves notable performance, demonstrating that SSPO does not require large amounts of data, and significant improvements in duration consistency can be achieved using a relatively small dataset.

\section{Conclusion}

In this study, we focus on the duration alignment task in video dubbing, which we consider as a preference optimization problem. To address this, we propose Segment Supervised Preference Optimization (SSPO) method. SSPO employs segment-wise sampling strategy and fine-grained preference alignment loss to mitigate the duration mismatch between source and target lines. Experiments demonstrate that SSPO achieves significant improvements in enhancing duration consistency between source and target speech compared to baseline methods, while maintaining translation quality.

\section*{Acknowledgments}

The authors would like to thank all the anonymous reviewers for their help and insightful comments.

\section*{Ethical Statement}

This research has been conducted with adherence to ethical guidelines and standards. The data utilized in the study, specifically the subtitles of film and television programs, were sourced from the Youku platform. All data collection and usage were performed following formal authorization and consent from Youku, ensuring that permissions were fully granted for academic research purposes.

The study respects intellectual property rights and confidentiality agreements, complying with all terms and conditions as stipulated by Youku. No personal or sensitive information was gathered or used during this research. The focus of the study remains strictly on the linguistic and translational aspects of the subtitled content.

We maintain a commitment to transparency and ethical integrity in research, ensuring that all findings and methodologies are presented honestly and without misrepresentation. This research seeks to contribute valuable insights to the field of subtitle translation while upholding the highest ethical standards.

\section*{Limitations}

Limitations of this study are listed as follows:

\paragraph{Emotion Induced Duration Variability.}
SSPO measures duration consistency by referencing the duration of synthesized speech from open-source TTS services. However, the duration of real visual media speech may vary due to factors such as character emotion, suggesting that the metric for duration consistency could be further optimized in future research. 

\paragraph{Language Dependent Alignment Limits.}
There is an upper limit to optimization in DA tasks. As demonstrated in experiments involving Spanish, inherent language characteristics can prevent the complete resolution of duration inconsistency, even under optimal conditions.

\bibliography{main}

\appendix

\section{Experimental Details}
\label{sec:expdetail}

In this section, we primarily describe the main experimental settings adopted in this study. Unless certain experiments require specific hyperparameters, we employ consistent hyperparameters across all experiments to maintain consistency and fairness in experimental comparisons.

\subsection{Data Resources}
\label{sec:dataset}

In this study, we use the Polylingual Subtitle Corpus (PolySC) from 42 films and TV series (2021-2024) on the online video platform Youku. The corpus includes original Chinese subtitles and professionally translated English, Thai, and Spanish subtitles, used for \texttt{zh}$\Rightarrow$\texttt{en}, \texttt{zh}$\Rightarrow$\texttt{th}, \texttt{zh}$\Rightarrow$\texttt{es}, and \texttt{es}$\Rightarrow$\texttt{zh} subtitle translation. Chinese is a high-information-density language, Thai is medium-density, while English and Spanish are low-density languages. We set $n=35$. Each translation direction's dataset contains approximately 26,000 prompt-response pairs for LLM training. For each direction, 97\% of the data is used as the SFT Demonstration dataset, including both prompts and responses. The remaining 3\% serves as the DA Query dataset, retaining only prompts. Statistics for the PolySC dataset are presented in Table~\ref{tab:sta}.

\begin{table*}[!h]
\centering
\begin{tabular}{cccc}
\Xhline{1.0pt}
\rowcolor{gray!20}
\textbf{Statistic} & \texttt{zh}$\Rightarrow$\texttt{en} & \texttt{zh}$\Rightarrow$\texttt{th} & \texttt{zh}$\Leftrightarrow$\texttt{es}\\
\hline
\textbf{period} & \multicolumn{3}{c}{2021-2024} \\
\textbf{\# plays} & \multicolumn{3}{c}{42} \\
\textbf{\# lines} & \multicolumn{3}{c}{684625} \\
\textbf{total duration (h)} & \multicolumn{3}{c}{471.4} \\
\hdashline
\textbf{\# avg source token} & \multicolumn{3}{c}{6.19 (\texttt{zh})} \\
\textbf{\# avg target token} & 8.17 & 21.11 & 8.66 (\texttt{es}) \\
\hdashline
\textbf{avg source duration (s)} & \multicolumn{3}{c}{1.096 (\texttt{zh})} \\
\textbf{avg target duration (s)} & 1.314 & 1.336 & 1.581 (\texttt{es})\\
\Xhline{1.0pt}
\end{tabular}
\caption{Statistics of the datasets.}
\label{tab:sta}
\end{table*}

PolySC dataset encompasses a diverse range of programs, including a total of 46 live-action television programs from 2021-2024 (42 for the training set and 4 for the test set). These programs span various genres such as fantasy, period drama, romance, and comedy, and include both long-form Series and mini series. The complete program list for PolySC dataset can be found at \url{https://github.com/CcQunResearch/SSPO/blob/main/SSPOTraining/Playlist.md}.

We present examples of Chinese and English subtitles from the PolySC dataset (in .ass file format) in Table~\ref{tab:subtitle}. The "Start" and "End" columns identify the beginning and end times of the lines in the episode. The purpose of DA is to align the duration of the LLM's translation with that of the original line. It is important to note that in our experiments, we did not use the interval between the "Start" and "End" columns as the duration for the lines. This is because in most productions, human-translated subtitles set the start and end times of the translation to match those of the original text. We use \texttt{edge-tts} as a unified standard to measure the duration of both the original and translated lines.

\begin{table*}[!h]
\centering
\resizebox{\textwidth}{!}{
\begin{tabular}{ccl}
\Xhline{1.0pt}
\rowcolor{gray!20}
\textbf{Start} & \textbf{End} & \textbf{Text} \\
\hline
0:02:21.96 & 0:02:24.87 & \begin{CJK}{UTF8}{gkai}你去城北的铺子买一些安神的香料\end{CJK} \\
0:02:25.32 & 0:02:27.39 & \begin{CJK}{UTF8}{gkai}大帅最近睡的不踏实\end{CJK} \\
0:02:27.55 & 0:02:29.27 & \begin{CJK}{UTF8}{gkai}他可以焚香办公\end{CJK} \\
0:02:29.27 & 0:02:31.48 & \begin{CJK}{UTF8}{gkai}这样也可以安心养神一些\end{CJK} \\
0:02:32.80 & 0:02:33.83 & \begin{CJK}{UTF8}{gkai}好嘞少夫人\end{CJK} \\
0:02:36.52 & 0:02:37.52 & \begin{CJK}{UTF8}{gkai}小雨\end{CJK} \\
0:02:37.52 & 0:02:38.92 & \begin{CJK}{UTF8}{gkai}你去城南的花坊\end{CJK} \\
0:02:38.92 & 0:02:40.67 & \begin{CJK}{UTF8}{gkai}购置些新鲜的鲜花回来\end{CJK} \\
0:02:41.67 & 0:02:43.24 & \begin{CJK}{UTF8}{gkai}咱们这聂府啊\end{CJK} \\
0:02:43.43 & 0:02:44.87 & \begin{CJK}{UTF8}{gkai}整日沉闷得很\end{CJK} \\
\hdashline
0:02:22.00 & 0:02:24.91 & \texttt{Go to the shop in the north and buy some calming incense ingredients} \\
0:02:25.36 & 0:02:27.43 & \texttt{for the Grand Marshal. He has had trouble sleeping well lately.} \\
0:02:27.59 & 0:02:29.31 & \texttt{He can burn incense while working} \\
0:02:29.31 & 0:02:31.52 & \texttt{to calm his nerves.} \\
0:02:32.84 & 0:02:33.87 & \texttt{OK, Young Madam.} \\
0:02:36.56 & 0:02:37.56 & \texttt{Xiaoyu,}  \\
0:02:37.56 & 0:02:38.96 & \texttt{go to the flower shop south of the city} \\
0:02:38.96 & 0:02:40.71 & \texttt{and buy some fresh flowers.} \\
0:02:41.71 & 0:02:43.28 & \texttt{Our Nie manor} \\
0:02:43.47 & 0:02:44.91 & \texttt{is dreary every day} \\
\Xhline{1.0pt}
\end{tabular}
}
\caption{Examples of Polylingual subtitles.}
\label{tab:subtitle}
\end{table*}

\subsection{Main Setting}
\label{sec:hp}

We primarily use \texttt{PyTorch}\footnote{\url{https://github.com/pytorch/pytorch}} and \texttt{Transformers}\footnote{\url{https://github.com/huggingface/transformers}} library to implement our methods, while leveraging \texttt{DeepSpeed}\footnote{\url{https://github.com/microsoft/DeepSpeed}} for multi-GPU parallel training. Due to the limitations imposed by the GLM4 series models' left-padding feature for batch texts, we employ \texttt{LLaMA-Factory}\footnote{\url{https://github.com/hiyouga/LLaMA-Factory}} library for GLM4-related experiments. We re-implemented the AutoDubbing and VideoDubber methods, and utilize the paid APIs provided by OpenAI and Anthropic to obtain experimental results related to GPT-3.5-Turbo, GPT-4, and Claude 3.5 Sonnet. In our main experiments, ablation studies, and extended experiments, we strive to maintain consistent non-relevant hyperparameters to ensure fairness and consistency in comparisons. All critical hyperparameter settings are presented in Table~\ref{tab:hp}.

\begin{table*}[!h]
\centering
\begin{tabular}{cccl}
 \Xhline{1.0pt}
 \rowcolor{gray!20}
 \textbf{Type} & \textbf{Hyperparameter} & \textbf{Value} & \textbf{Remark} \\
 \hline
 \multirow{6}{*}{\textbf{Sampling}} & $\varepsilon _{1}$ & 4 & \multirow{2}{*}{segment-level sampling indicator function threshold} \\
 ~ & $\varepsilon _{2}$ & 0.08 & ~ \\
 ~ & $k$ & 20 & sampling number \\
 ~ & \textbf{temperature} & 1.4 & \multirow{3}{*}{sampling text generation hyperparameters} \\
 ~ & \textbf{top k} & 60 & ~ \\
 ~ & \textbf{top p} & 0.95 & ~ \\
 \hdashline
 \multirow{5}{*}{\textbf{Training}} & \textbf{optimizer} & AdamW & - \\
 ~ & \textbf{learning rate} & 4e-6 & - \\
 ~ & \textbf{epoch} & 4 & - \\
 ~ & \textbf{batch size} & 64 & \# lines \\
 ~ & \textbf{total data size} & 1e4 & \# training lines \\
 \hdashline
 \multirow{3}{*}{\textbf{LoRA}} & \textbf{lora $r$} & 16 & - \\
 ~ & \textbf{lora $\alpha$} & 32 & - \\
 ~ & \textbf{lora targets} & Q\&K\&V & - \\
 \hdashline
 \textbf{TKLD} & $\lambda$ & 1e-4 & weight of the TKLD divergence constraint \\
 \hdashline
 \textbf{SSPO} & $\beta$ & 0.5 & hyperparameter in SSPO loss \\
 \Xhline{1.0pt}
\end{tabular}
\caption{Hyperparameter configuration.}
\label{tab:hp}
\end{table*}

\subsection{Computational Resources}

We conduct all experiments on 8 A800 80GB SXM GPUs. The time required for the three stages - SFT, sampling, and SSPO Training - is approximately 2h, 3h, and 1.5h, respectively. It's worth noting that due to the non-parallelizable nature of LLM inference, the sampling stage is more time-consuming compared to the training stage. Overall, executing a complete workflow on 8 A800 GPUs can be accomplished within an acceptable time frame.

\section{Extended Experiments}
\label{sec:extendedexp}

In this section, we will present additional evaluation experiments of SSPO across various aspects.

\subsection{Performance on Bidirectional Translation}
\label{sec:expes}

In this subsection, we validate the performance of SSPO on \texttt{zh}$\Rightarrow$\texttt{es} and \texttt{es}$\Rightarrow$\texttt{zh} translations.

\subsubsection{Experiments}

We conducted experiments on \texttt{zh}$\Rightarrow$\texttt{es} and \texttt{es}$\Rightarrow$\texttt{zh} translation tasks, with the results presented in Table~\ref{tab:exp-es}. The findings demonstrate that SSPO consistently improves the duration consistency of the translated content. When translating from Chinese, a language with high information density, to Spanish, which has lower information density, SSPO reduces the proportion of translated lines exceeding the duration of the original lines. Conversely, when translating from Spanish to Chinese, SSPO decreases the proportion of original lines exceeding the duration of the translated lines. Simultaneously, the average duration overrun is reduced. These experiments on Spanish translation further validate the universal effectiveness of SSPO in addressing duration inconsistencies arising from disparities in language information density.

\begin{table*}[!h]
\centering
\resizebox{\textwidth}{!}{
\begin{tabular}{cccccccc|cccccc}
 \Xhline{1.0pt}
 \rowcolor{gray!20}
 ~ & ~ & \multicolumn{6}{c}{\textbf{\texttt{zh}$\Rightarrow$\texttt{es}}} & \multicolumn{6}{c}{\textbf{\texttt{es}$\Rightarrow$\texttt{zh}}}\\
 \cline{3-14}
 \rowcolor{gray!20}
 \multirow{-2}{*}{\textbf{Method}} & \multirow{-2}{*}{\textbf{Train}} & \textbf{S>T Rate} & \textbf{S>T Dur} & \textbf{T>S Rate} & \textbf{T>S Dur} & \textbf{CR} & $\mathcal{P}$ & \textbf{S>T Rate} & \textbf{S>T Dur} & \textbf{T>S Rate} & \textbf{T>S Dur} & \textbf{CR} & $\mathcal{P}$\\
 \hline
 Gold Reference & - & \textit{6.2\%} & \textit{0.356} & \textit{86.0\%} & \textit{0.801} & \textit{7.8\%} & \textit{1.490} & \textit{86.2\%} & \textit{0.808} & \textit{6.2\%} & \textit{0.355} & \textit{7.6\%} & \textit{0.590} \\
 \hdashline
 AutoDubbing & SFT & 4.3\% & \colorbox{mlo}{\textbf{0.239}} & 91.3\% & 0.744 & 4.4\% & 1.304 & 92.1\% & 0.757 & \colorbox{mlo}{\textbf{1.9\%}} & 0.237 & 6.0\% & 0.564 \\
 VideoDubber & SFT & 5.3\% & 0.254 & 88.3\% & 0.732 & 6.4\% & 1.273 & 87.4\% & 0.731 & 5.7\% & \colorbox{mlo}{\textbf{0.223}} & 6.9\% & 0.535 \\ 
 \hdashline
 GPT-3.5-Turbo & PE & \colorbox{mlb}{\textbf{1.6\%}} & \colorbox{mlb}{\textbf{0.223}} & 94.9\% & 0.937 & 3.5\% & 2.124 & 94.9\% & 0.941 & \colorbox{mlb}{\textbf{1.8\%}} & \colorbox{mlb}{\textbf{0.222}} & 3.3\% & 0.720 \\
 GPT-4o & PE & 3.0\% & 0.252 & 90.6\% & 0.779 & 6.3\% & 1.429 & 91.2\% & 0.791 & 2.8\% & 0.257 & 5.9\% & 0.589 \\
 Claude 3.5 Sonnet & PE & \colorbox{mlo}{\textbf{2.3\%}} & 0.242 & 92.3\% & 0.756 & 5.4\% & 1.290 & 92.6\% & 0.760 & \colorbox{mlo}{\textbf{1.9\%}} & 0.246 & 5.5\% & 0.570 \\
 \hdashline
 \multirow{2}{*}{Llama3.1-8B-CN-Chat} & SFT & 5.8\% & 0.336 & 86.3\% & 0.793 & 7.9\% & 1.502 & 91.8\% & 0.826 & 2.6\% & 1.223 & 5.7\% & 0.620 \\
 ~ & \textbf{SSPO} & 8.2\% & 0.315 & 80.3\% & 0.614 & 11.5\% & 0.858 & \colorbox{mlo}{\textbf{82.2\%}} & \colorbox{mlo}{\textbf{0.640}} & 6.9\% & 0.483 & \colorbox{mlo}{\textbf{10.9\%}} & \colorbox{mlo}{\textbf{0.450}} \\
 \hdashline
 \multirow{2}{*}{GLM-4-9B-Chat} & SFT & 5.9\% & 0.346 & 85.8\% & 0.785 & 8.3\% & 1.492 & 91.9\% & 0.834 & 2.5\% & 0.601 & 5.6\% & 0.624 \\
 ~ & \textbf{SSPO} & 15.9\% & 0.362 & \colorbox{mlo}{\textbf{70.2\%}} & \colorbox{mlo}{\textbf{0.548}} & \colorbox{mlo}{\textbf{13.9\%}} & \colorbox{mlo}{\textbf{0.665}} & \colorbox{mlb}{\textbf{81.5\%}} & 0.654 & 7.5\% & 0.542 & \colorbox{mlb}{\textbf{11.0\%}} & 0.460 \\
 \hdashline
 \multirow{2}{*}{Qwen2.5-14B-Instruct} & SFT & 7.0\% & 0.335 & 84.0\% & 0.758 & 8.9\% & 1.320 & 91.8\% & 0.830 & 2.6\% & 0.297 & 5.6\% & 0.622 \\
 ~ & \textbf{SSPO} & 15.8\% & 0.359 & \colorbox{mlb}{\textbf{69.6\%}} & \colorbox{mlb}{\textbf{0.538}} & \colorbox{mlb}{\textbf{14.6\%}} & \colorbox{mlb}{\textbf{0.644}} & 84.8\% & \colorbox{mlb}{\textbf{0.629}} & 4.8\% & 0.308 & 10.3\% & \colorbox{mlb}{\textbf{0.447}} \\
 \hdashline
 Alignment Bound & - & \textit{4.3\%} & \textit{0.291} & \textit{84.3\%} & \textit{0.566} & \textit{11.3\%} & \textit{0.654} & \textit{86.2\%} & \textit{0.660} & \textit{1.4\%} & \textit{0.245} & \textit{12.4\%} & \textit{0.466} \\
 \Xhline{1.0pt}
\end{tabular}
}
\caption{\texttt{zh}$\Rightarrow$\texttt{es} and \texttt{es}$\Rightarrow$\texttt{zh} results on test set. The best and second best results are denoted as \colorbox{mlb}{\textbf{blue}} and \colorbox{mlo}{\textbf{orange}}.}
\label{tab:exp-es}
\end{table*}

\subsubsection{Visualization}

\begin{figure*}[!h]
  \centering
  \subfigure[\texttt{zh}$\Rightarrow$\texttt{es}]{\includegraphics[width=0.48\textwidth]{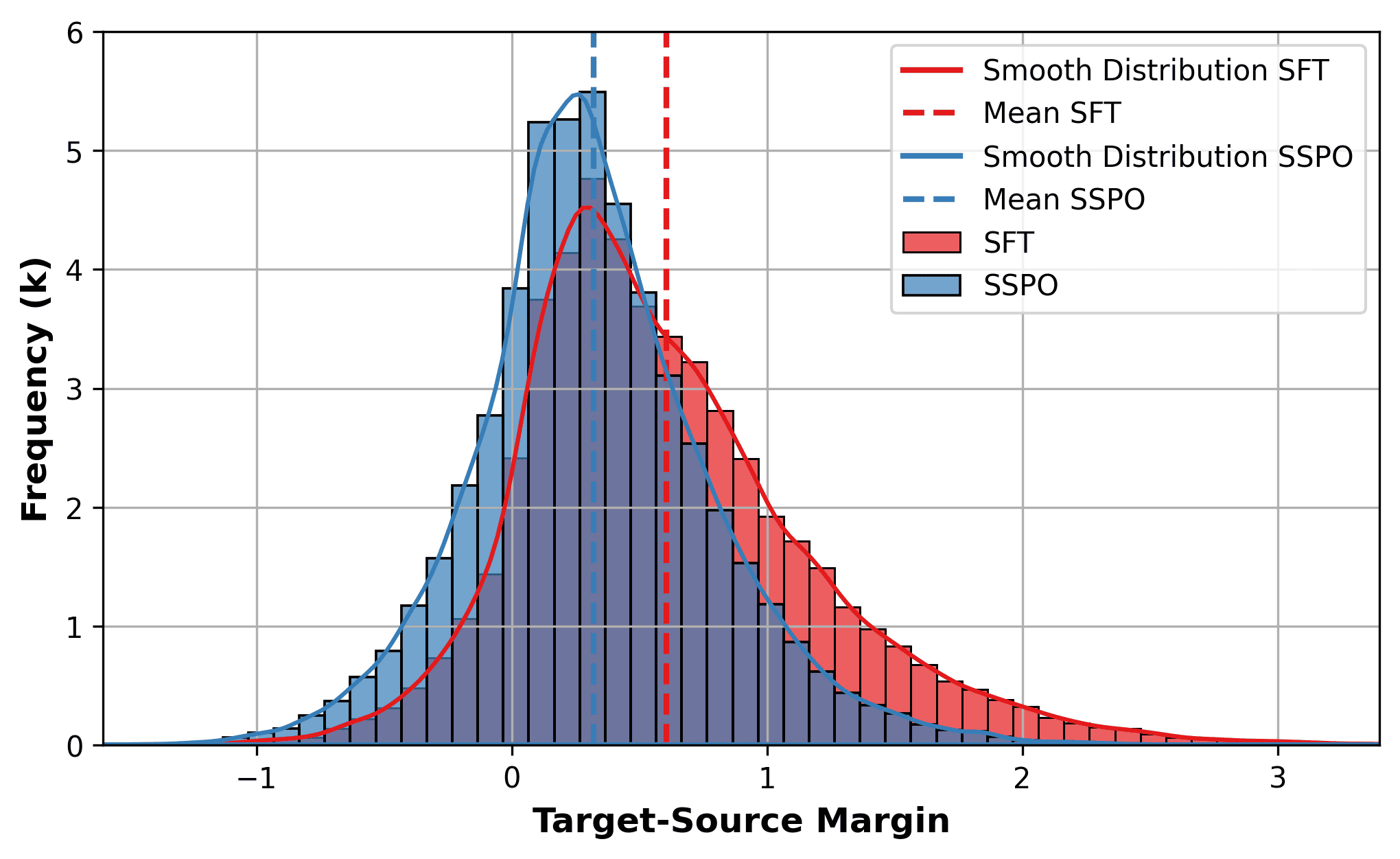}}
  \subfigure[\texttt{es}$\Rightarrow$\texttt{zh}]{\includegraphics[width=0.48\textwidth]{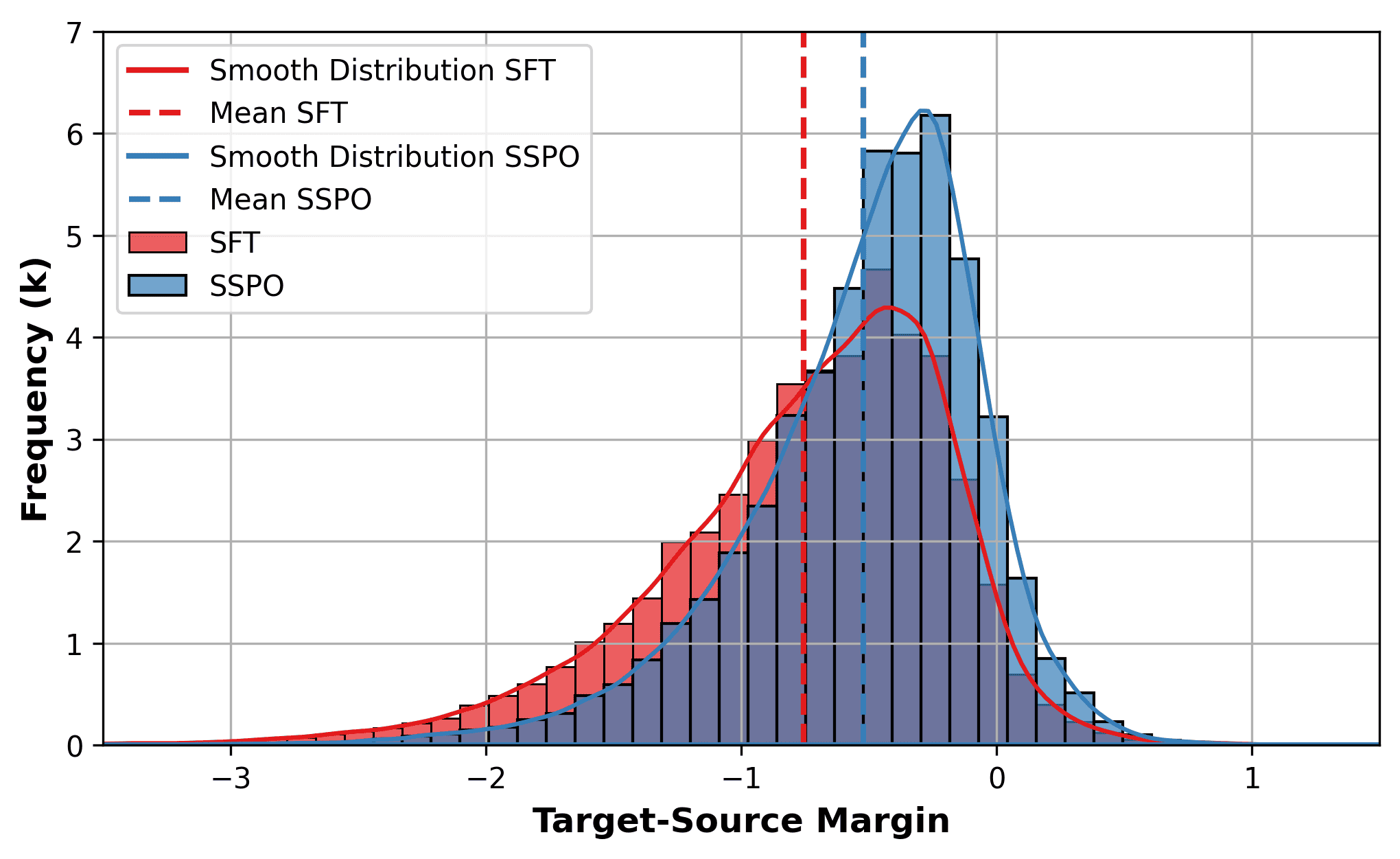}}
  \caption{Frequency distribution of Qwen2.5-14B-Instruct model on \texttt{zh}$\Rightarrow$\texttt{es} and \texttt{es}$\Rightarrow$\texttt{zh} translations.}
  \label{fig:vises}
\end{figure*}

In Figure~\ref{fig:vises}, we present the frequency distribution of the duration differences between the translated and original content for both the SFT and SSPO models of Qwen2.5-14B-Instruct. Similar to our previous experiments, it is evident that after SSPO training, the duration discrepancies between the original and translated content are significantly reduced. The histogram for the SSPO model is noticeably more concentrated around zero compared to that of the SFT model. This observation indicates that regardless of the information density disparities between the source and target languages, SSPO consistently improves duration consistency, resulting in a shift of the histogram towards zero.

\subsection{Further Exploration}
\label{sec:furtherexp}

We explore two alternative solutions here.

\subsubsection{Vanilla DPO Training}

In considering DA as a preference optimization problem, we sought to investigate the question: "Can the vanilla DPO algorithm directly solve the DA problem?" Based on this inquiry, we conducted relevant designs and experiments. SSPO achieves good control over the duration of translated lines through a segment-wise sampling strategy and fine-grained DPO loss. We aimed to validate the impact of these two components. We employed the standard DPO training process \cite{17} to perform DA on the SFT model. Specifically, we utilized either coarse-grained or fine-grained sampling to sample a chosen response $y^{(\text{c})}$ and a rejected response $y^{(\text{r})}$ for each sample $x\in \mathcal{D}_{\text{query}}$ in the Query dataset. Subsequently, we optimized the policy model using the standard DPO loss \cite{17}.

The adopted coarse-grained and fine-grained sampling procedures are illustrated in Algorithm~\ref{alg:dpocoarse} and Algorithm~\ref{alg:dpofine}, respectively. In coarse-grained sampling, $k$ complete responses are directly sampled for a given prompt $x$. The consistency penalty $\mathcal{P}(s_i,t_i)$ is calculated for each of the $n$ lines in each response, and the sum of these penalties is computed. The response with the minimum sum is selected as the chosen response $y^{(\text{c})}$, while the one with the maximum sum becomes the rejected response $y^{(\text{r})}$. Fine-grained sampling, on the other hand, requires two segment-wise sampling cycles similar to those in Algorithm~\ref{alg:sampling} for a single prompt $x$. In these two cycles, the lines with the minimum and maximum $\mathcal{P}$ are used as the prefix for sampling the next line, respectively. This process ultimately yields the segment-wise sampled chosen response $y^{(\text{c})}$ and rejected response $y^{(\text{r})}$.

\begin{algorithm*}[p]
  \caption{Coarse-grained Sampling for Vanilla DPO.}
  \label{alg:dpocoarse}
  \begin{algorithmic}[1]
    \Require SFT model $\pi _{\text{sft}}$, query dataset $\mathcal{D}_{\text{query}}$, sampling number $k$.
    \Ensure sampled response pairs set $\mathcal{S}(x)$.
    \For {any $x\in \mathcal{D}_{\text{query}}$}
    \State $//$ Sample multiple candidate responses.
    \For {$i=1$ to $k$}
    \State Sample $\pi _{\text{sft}}(y|x)$.
    \EndFor
    \State Measure the sum of $\mathcal{P}$ for each line of $y^{i}$ in the candidate set $\left \{y^{i}|i=1,2,\dots ,k\right \}$.
    \State Select chosen $y^{(\text{c})}$ and rejected $y^{(\text{r})}$.
    \EndFor \\
    \Return $\mathcal{S}(x)\equiv \left \{y^{(\text{c})},y^{(\text{r})}\right \}$.
  \end{algorithmic}
\end{algorithm*}

\begin{algorithm*}[p]
  \caption{Fine-grained Sampling for Vanilla DPO.}
  \label{alg:dpofine}
  \begin{algorithmic}[1]
    \Require SFT model $\pi _{\text{sft}}$, query dataset $\mathcal{D}_{\text{query}}$, sampling number $k$.
    \Ensure sampled response pairs set $\mathcal{S}(x)$.
    \For {any $x\in \mathcal{D}_{\text{query}}$}
    \State $//$ The first sampling cycle used to obtain $y^{(\text{c})}$.
    \For {$i=1$ to $n$}
    \For {$j=1$ to $k$}
    \State Sample $\pi _{\text{sft}}(t_{i}^{j}|x,s_{1},t_{1}^{\text{(c)}},\dots ,s_{i-1},t_{i-1}^{\text{(c)}},s_{i})$.
    \EndFor
    \State Deduplicate and measure $\{t_{i}^{j}|j=1,2,\dots ,k\}$ by $\mathcal{P}(s_i,t_i^j)$, and select chosen $t_{i}^{\text{(c)}}$.
    \EndFor 
    \State Concatenate $\{(s_{i},t_{i}^{\text{(c)}})|i=1,2,\dots ,n\}$ yields $y^{(\text{c})}$.
    \State $//$ The second sampling cycle used to obtain $y^{(\text{r})}$.
    \For {$i=1$ to $n$}
    \For {$j=1$ to $k$}
    \State Sample $\pi _{\text{sft}}(t_{i}^{j}|x,s_{1},t_{1}^{\text{(r)}},\dots ,s_{i-1},t_{i-1}^{\text{(r)}},s_{i})$.
    \EndFor
    \State Deduplicate and measure $\{t_{i}^{j}|j=1,2,\dots ,k\}$ by $\mathcal{P}(s_i,t_i^j)$, and select rejected $t_{i}^{\text{(r)}}$.
    \EndFor 
    \State Concatenate $\{(s_{i},t_{i}^{\text{(r)}})|i=1,2,\dots ,n\}$ yields $y^{(\text{r})}$. 
    \EndFor \\
    \Return $\mathcal{S}(x)\equiv \{y^{(\text{c})},y^{(\text{r})}\}$.
  \end{algorithmic}
\end{algorithm*}

\subsubsection{Advantage-based PPO Training}

\begin{algorithm*}[p]
  \caption{Sampling for PPO solution.}
  \label{alg:ppo}
  \begin{algorithmic}[1]
    \Require SFT model $\pi _{\text{sft}}$, DA dataset $\mathcal{D}_{\text{query}}$.
    \Ensure sampled trajectory set $\mathcal{T}\equiv \{\tau \}$.
    \For {any $x\in \mathcal{D}_{\text{query}}$}
    \State $//$ Iterate through the dialogue lines in $x$.
    \For {$i=1$ to $n$}
    \State Sample $\pi _{\text{sft}}(t_{i}|x,s_{1},t_{1},\dots ,s_{i-1},t_{i-1},s_{i})$ and measure $t_{i}$ by $\mathcal{P}(s_i,t_i)$.
    \EndFor
    \State Obtain trajectory $\tau =\{(p_i,t_i,\mathcal{P}(s_i,t_i))|i=1,2,\cdots ,n\}$.
    \EndFor \\
    \Return $\mathcal{T}\equiv \{\tau \}$.
  \end{algorithmic}
\end{algorithm*}

\begin{table*}[t]
\centering
\resizebox{\textwidth}{!}{
\begin{tabular}{ccccccccc|cccccc}
 \Xhline{1.0pt}
 \rowcolor{gray!20}
 ~ & ~ & ~ & \multicolumn{6}{c}{\textbf{\texttt{zh}$\Rightarrow$\texttt{en}}} & \multicolumn{6}{c}{\textbf{\texttt{zh}$\Rightarrow$\texttt{th}}}\\
 \cline{4-15}
 \rowcolor{gray!20}
 \multirow{-2}{*}{\textbf{Method}} & \multirow{-2}{*}{\textbf{Training}}  & \multirow{-2}{*}{\textbf{Sampling}} & \textbf{S>T Rate} & \textbf{S>T Dur} & \textbf{T>S Rate} & \textbf{T>S Dur} & \textbf{CR} & $\mathcal{P}$ & \textbf{S>T Rate} & \textbf{S>T Dur} & \textbf{T>S Rate} & \textbf{T>S Dur} & \textbf{CR} & $\mathcal{P}$\\
 \hline
 Gold Reference & - & - & 18.0\% & 0.344 & 64.1\% & 0.464 & 17.9\% & 0.501 & 19.4\% & 0.369 & 60.2\% & 0.460 & 20.3\% & 0.489 \\
 \hdashline
 \multirow{5}{*}{GLM-4-9B-Chat} & SFT & - & \colorbox{mlo}{\textbf{19.5\%}} & \colorbox{mlo}{\textbf{0.342}} & 60.5\% & 0.427 & 20.0\% & 0.428 & \colorbox{mlb}{\textbf{18.1\%}} & \colorbox{mlb}{\textbf{0.291}} & 55.0\% & 0.391 & 27.0\% & 0.360 \\
 ~ & \multirow{2}{*}{DPO} & C & \colorbox{mlb}{\textbf{19.4\%}} & \colorbox{mlb}{\textbf{0.340}} & 61.1\% & 0.432 & 19.4\% & 0.432 & 18.7\% & 0.293 & 54.4\% & 0.391 & 26.9\% & 0.358 \\
 ~ & ~ & F & \colorbox{mlo}{\textbf{19.5\%}} & 0.343 & 60.7\% & 0.424 & 19.8\% & 0.431 & \colorbox{mlo}{\textbf{18.4\%}} & 0.289 & 55.1\% & 0.392 & 26.6\% & 0.360  \\
 ~ & PPO & F & 24.5\% & \colorbox{mlo}{\textbf{0.342}} & \colorbox{mlo}{\textbf{53.4\%}} & \colorbox{mlo}{\textbf{0.364}} & \colorbox{mlo}{\textbf{22.1\%}} & \colorbox{mlo}{\textbf{0.323}} & 23.5\% & 0.290 & \colorbox{mlo}{\textbf{47.5\%}} & \colorbox{mlo}{\textbf{0.318}} & \colorbox{mlo}{\textbf{29.0\%}} & \colorbox{mlo}{\textbf{0.301}} \\
 ~ & \textbf{SSPO} & F & 29.6\% & 0.350 & \colorbox{mlb}{\textbf{45.9\%}} & \colorbox{mlb}{\textbf{0.323}} & \colorbox{mlb}{\textbf{24.5\%}} & \colorbox{mlb}{\textbf{0.283}} & 25.9\% & \colorbox{mlb}{\textbf{0.291}} & \colorbox{mlb}{\textbf{42.0\%}} & \colorbox{mlb}{\textbf{0.318}} & \colorbox{mlb}{\textbf{32.1\%}} & \colorbox{mlb}{\textbf{0.254}} \\
 \hdashline
 \multirow{5}{*}{Qwen2.5-14B-Instruct} & SFT & - & \colorbox{mlb}{\textbf{20.0\%}} & \colorbox{mlb}{\textbf{0.341}} & 59.8\% & 0.439 & 20.2\% & 0.423 & 18.2\% & 0.294 & 55.4\% & 0.397 & 26.4\% & 0.362 \\
 ~ & \multirow{2}{*}{DPO} & C & \colorbox{mlo}{\textbf{25.4\%}} & 0.376 & 54.0\% & 0.508 & 20.6\% & 0.408 & \colorbox{mlo}{\textbf{18.1\%}} & 0.297 & 55.9\% & 0.411 & 25.9\% & 0.369 \\
 ~ & ~ & F & 25.6\% & 0.378 & 54.2\% & 0.492 & 20.2\% & 0.408 & \colorbox{mlb}{\textbf{17.7\%}} & 0.295 & 56.4\% & 0.411 & 26.0\% & 0.373 \\
 ~ & PPO & F & 30.2\% & 0.372 & \colorbox{mlo}{\textbf{47.4\%}} & \colorbox{mlo}{\textbf{0.366}} & \colorbox{mlo}{\textbf{22.4\%}} & \colorbox{mlo}{\textbf{0.315}} & 30.8\% & \colorbox{mlo}{\textbf{0.293}} & \colorbox{mlo}{\textbf{38.3\%}} & \colorbox{mlo}{\textbf{0.314}} & \colorbox{mlo}{\textbf{30.9\%}} & \colorbox{mlo}{\textbf{0.237}} \\
 ~ & \textbf{SSPO} & F & 34.4\% & \colorbox{mlo}{\textbf{0.366}} & \colorbox{mlb}{\textbf{40.6\%}} & \colorbox{mlb}{\textbf{0.324}} & \colorbox{mlb}{\textbf{24.9\%}} & \colorbox{mlb}{\textbf{0.272}} & 38.6\% & \colorbox{mlb}{\textbf{0.290}} & \colorbox{mlb}{\textbf{25.3\%}} & \colorbox{mlb}{\textbf{0.279}} & \colorbox{mlb}{\textbf{36.1\%}} & \colorbox{mlb}{\textbf{0.198}} \\
 \hdashline
 Alignment Bound & - & - & \textit{16.4\%} & \textit{0.278} & \textit{39.3\%} & \textit{0.331} & \textit{44.3\%} & \textit{0.220} & \textit{9.2\%} & \textit{0.232} & \textit{40.4\%} & \textit{0.313} & \textit{50.4\%} & \textit{0.203} \\
 \Xhline{1.0pt}
\end{tabular}
}
\caption{Experimental evaluation results of alternative solutions. C for Coarse-grained, while F for Fine-grained. The best and second best results are denoted as \colorbox{mlb}{\textbf{blue}} and \colorbox{mlo}{\textbf{orange}}.}
\label{tab:further-exp}
\end{table*}

In the literature, RLHF (primarily based on Proximal Policy Optimization (PPO)) is often considered to outperform DPO\cite{po1,po2} despite its complexity. Thus, we want to explore the question, "Can PPO techniques achieve better results than DPO-based SSPO in DA?" Based on this hypothesis, we conducted related experiments. We employed an advantage-based PPO training process\cite{16-2}. Specifically, we implement the PPO training process through the following steps:

\begin{enumerate}[label=\arabic*.]
\item \textit{Rollout} - For each sample $x\in \mathcal{D}_{\text{query}}$ in query dataset, sample a trajectory $\tau$ using Algorithm~\ref{alg:ppo}.
\item \textit{Compute Rewards and Advantages} - Perform Generalized Advantage Estimation (GAE) on $\tau$ using the value network $V_{\phi}$ (with Qwen2.5-7B-Instruct \cite{qwen25} as the backbone):
\begin{equation}
\delta _{i}=-\mathcal{P}(s_i,t_i)+\gamma V_{\phi }(p_{i+1})-V_{\phi }(p_{i}),
\end{equation}
\begin{equation}
A_{i}=\displaystyle\sum_{l=0}^{n-i-1}(\gamma \lambda )^{l}\delta _{i+l}.
\end{equation}
\item \textit{Update Policy Network} - Let $\pi _{\text{old}}$ be the fixed old policy during sampling, and minimize the loss function (the KL divergence constraint between $\pi _{\theta}$ and $\pi _{\text{old}}$ is omitted here):
\begin{equation}
\begin{split}
&\mathcal{L}_{\text{clip}}(\theta )=\\&-\mathbb{E}_{x\sim \mathcal{D}_{\text{query}}} \Bigg [ \sum_{i=1}^{n} \min \Bigg ( \frac{\pi _{\theta }(t_i\mid p_i)}{\pi _{\text{old}}(t_i\mid p_i)}A_i,\\&\text{clip}\bigg (\frac{\pi _{\theta }(t_i\mid p_i)}{\pi _{\text{old}}(t_i\mid p_i)},1-\epsilon ,1+\epsilon \bigg ) A_i \Bigg ) \Bigg].
\end{split}
\end{equation}
\item \textit{Update Value Network} - Fit $V_{\phi}(p_{i})$ to the GAE-based estimated value $\hat{V}_{i}=A_{i}+V_{\phi}(p_{i})$ using mean squared error:
\begin{equation}
\mathcal{L}_{V}(\phi )=\mathbb{E}_{x\sim \mathcal{D}_{\text{query}}}\left [\displaystyle\sum_{i=1}^{n}(V_{\phi }(p_{i})-\hat{V}_{i})^{2}\right ].
\end{equation}
\item \textit{Iterate Multiple Rounds} - Continuously collect new data and update the value network and policy model until convergence.
\end{enumerate}

\subsubsection{Evaluation}

We utilized the GLM-4-9B-Chat and Qwen2.5-14B-Instruct backbone models to validate the DA performance in \texttt{zh}$\Rightarrow$\texttt{en} and \texttt{zh}$\Rightarrow$\texttt{th} translations under the "Vanilla DPO Training" and "Advantage-based PPO Training" configurations. The experimental results are presented in Table~\ref{tab:further-exp}. The results indicate that while PPO training achieved a performance improvement over the SFT model, none of the other configurations showed significant enhancement, and their performance is notably different from that of SSPO. With vanilla DPO training, neither coarse-grained nor fine-grained sampling improved duration consistency, and the fine-grained sampling it employed consumed twice the time and computational resources compared to SSPO. This validates that the effectiveness of SSPO arises from the dual factors of sentence-level sampling strategy and fine-grained DPO loss. This means that each sentence in the prompt must be sampled and optimized independently, as sampling and loss calculation on the complete response are ineffective. Additionally, although RLHF methods often outperform DPO methods in LLM preference alignment tasks, using PPO methods in DA tasks resulted in performance lower than SSPO. This is because, for DA tasks, duration consistency has a clear metric (i.e., $\mathcal{P}$), and thus SSPO optimizes towards the optimal solution by increasing the generation probability of the most consistent translations for each sentence and reducing the generation probability of the least consistent translations, while PPO training follows a gradual optimization process. This contrasts with the situation in preference alignment \cite{po1,po2}.

\section{Theory: Segment Supervised Preference Optimization}
\label{sec:theory}

In this section, we formalize the localized multi-segment preference optimization problem and validate the effectiveness of SSPO, while also highlighting the limitations of general preference optimization methods. Note that the notation used in this section may have slightly different meanings from those in previous sections.

\subsection{Localized Multi-Segment Preference Optimization Problem}

For a language model $\pi _{\theta }$, given an input $x \in \mathcal{X}$ (where $\mathcal{X}$ is the input space), assume it consists of $n$ interrelated segments, i.e., $x = (x_1, x_2, \ldots, x_n)$. Correspondingly, the output $y \in \mathcal{Y}$ (where $\mathcal{Y}$ is the output space) also comprises $n$ interrelated segments, i.e., $y = (y_1, y_2, \ldots, y_n)$, with each $x_i$ corresponding to $y_i$. Additionally, the generation of $y_i$ is influenced by $y_1, y_2, \ldots, y_{i-1}$ (note that this influence may not only stem from the autoregressive property of the language model but also from semantic dependencies among output segments), expressed as:
\begin{equation}
\pi _{\theta }(y\mid x)=\displaystyle\prod_{i=1}^{n}\pi _{\theta }(y_{i}\mid x,y_1,\cdots ,y_{i-1}).
\end{equation}

The general preference optimization task involves an outcome-supervised reward function $r(x,y)$ for the output $y$ given input $x$. In contrast, the localized multi-segment preference optimization problem employs segment-supervised preference metrics, denoted as $r(x_i,y_i)$, which quantify the alignment of segment output $y_i$ with the predefined optimization preference for its corresponding segment input $x_i$. Our objective is to adjust model parameters $\theta$ during training such that the policy $\pi_{\theta}$ is optimized to maximize $r(x_i,y_i)$ across all segments.

\subsection{Ineffectiveness of General Preference Optimization Methods}

Let's take DPO \cite{17} as an example to illustrate the limitations of general preference optimization methods when dealing with localized multi-segment preference optimization problems. In conducting localized preference optimization, DPO first labels the complete preferred response $y^w = (y_1^w, y_2^w, \ldots, y_n^w)$ and the less preferred response $y^l = (y_1^l, y_2^l, \ldots, y_n^l)$ corresponding to $x = (x_1, x_2, \ldots, x_n)$ using $r(x_i, y_i)$. $y^w$ and $y^l$ necessarily satisfy:
\begin{equation}
\begin{split}
&\pi _{\theta }(y^w\mid x)=\displaystyle\prod_{i=1}^{n}\pi _{\theta }(y_i^w\mid x,y_1^w,\cdots ,y_{i-1}^w),\\&\pi _{\theta }(y^l\mid x)=\displaystyle\prod_{i=1}^{n}\pi _{\theta }(y_i^l\mid x,y_1^l,\cdots ,y_{i-1}^l).
\end{split}
\end{equation}

Then, DPO optimizes $\pi _{\theta }$ by increasing the log probability of the preferred response relative to the less preferred response. For the contrastive term in DPO loss, we calculate:
\begin{equation}
\resizebox{0.48\textwidth}{!}{$
\begin{aligned}
&\text{log}\frac{\pi _{\theta }(y^{w} \mid x)}{\pi _{\text{ref}}(y^{w} \mid x)} - \text{log}\frac{\pi _{\theta }(y^{l} \mid x)}{\pi _{\text{ref}}(y^{l} \mid x)} =\\& \displaystyle\sum_{i=1}^{n}\left [\text{log}\frac{\pi _{\theta }(y_i^w \mid x,y_{1:i-1}^w)}{\pi _{\text{ref}}(y_i^w \mid x,y_{1:i-1}^w)} - \text{log}\frac{\pi _{\theta }(y_i^l \mid x,y_{1:i-1}^l)}{\pi _{\text{ref}}(y_i^l \mid x,y_{1:i-1}^l)}\right ],
\end{aligned}
$}
\end{equation}
where $y_{1:i-1}^w = (y_1^w, \ldots, y_{i-1}^w)$ and $y_{1:i-1}^l = (y_1^l, \ldots, y_{i-1}^l)$. 

It can be observed that the generation probability of the $i$-th segment is not conditioned on the same prefix, but rather on its own preferred or dispreferred prefix (i.e., $y_{1:i-1}^w$ or $y_{1:i-1}^l$). As a result, the comparison of the $i$-th segment is not a fair "apples-to-apples" comparison. DPO only performs a single holistic preference judgment on the complete sequences $(y^w, y^l)$, leaving the model unaware of how to adjust each individual segment. In other words, while the model knows that the full sequence $y^w$ is superior to $y^l$, it lacks guidance on how to make locally optimal choices for the $i$-th segment. Thus, as demonstrated by the experimental results in Appendix~\ref{sec:furtherexp}, vanilla DPO fails to effectively address the task of localized multi-segment preference optimization.

\subsection{Segment Supervised Preference Optimization}

Unlike DPO, SSPO labels the preferred response $y_i^w$ and dispreferred response $y_i^l$ for the next segment individually, using $y_{1:i-1}^w$ as the prefix, instead of obtaining the entire response sequence. Specifically, for the $i$-th segment, the prefix is the fixed prefix $p_i=(x,y_1^w,\cdots ,y_{i-1}^w)$. This ensures that each segment is compared on the same and preferred prefix, eliminating unfair competition between `preferred prefix vs. dispreferred prefix'. Under the same prefix, preference alignment loss (such as that used in DPO) is applied to compare $\pi _{\theta }(y_i^w \mid p_i)$ and $\pi _{\theta }(y_i^l \mid p_i)$ conditioned on the same $p_i$. Ultimately, the cumulative loss of all segments enforces preference constraints on each segment:
\begin{equation}
\begin{split}
&\mathcal{L}_{\text{SSPO}}(\pi _{\theta };\pi _{\text{ref}})=\\&-\mathbb{E}_{(x,y_{1:n}^{w},y_{1:n}^{l})\sim \mathcal{D}}\left [\displaystyle\sum_{i=1}^{n}\mathcal{L}_{\text{DPO}}(y_i^{w},y_i^{l},p_i)\right ].
\end{split}
\end{equation}

\section{Discussion}
\label{sec:discussion}

In this section, we will present further discussions on the SSPO method.

\subsection{Recommendations}

We offer the following development suggestions for technicians using SSPO:

\begin{itemize}
\item Generally, larger models will consume more time and computational resources during the sampling phase, necessitating a trade-off between performance and cost.
\item Due to the varying token encoding densities for different languages in LLM, a larger number of sampling tokens should be set for Thai (e.g., 80).
\item For languages with similar information density (e.g., English and German), where duration consistency is not critical or the TTS stage is not required, DA may be omitted.
\item When applying SSPO to other localized preference optimization tasks, it is essential to determine the task's preference metrics and criteria for optimization exemptions.
\end{itemize}

\subsection{Future Research}

Translation quality evaluation experiments indicate that LLMs' translations are inferior to human translations in terms of vividness. The subtitle texts of visual media programs are typically deeply integrated with their associated video and audio. Compared to the translation of legal and religious texts, subtitle translation may not require strict accuracy; instead, it should focus more on the vividness of the translation. In future research, we aim to apply SSPO to improve subtitle translation quality, exploring methods and techniques to enhance the vividness of translations.

\section{Prompt and Instructions}
\label{sec:pi}

In this section, we present the input and output formats of the LLMs we employed, as well as the evaluation instructions used for manual assessment of translation quality.

\subsection{Input and Output of LLM}
\label{sec:io}

We present the prompt and response formats for the \texttt{zh}$\Rightarrow$\texttt{en} SFT translation model in Table~\ref{tab:prompt} and Table~\ref{tab:response} ( (similar for other languages)). We process the original subtitle text and its translation of the television programs into this format for training the SFT model. The prompt is structured as follows: 

\begin{enumerate}[label=\arabic*.]
\item \textit{Preamble} - An introduction and instructions describing the task at hand
\item \textit{Terminology} - A translation glossary for terminology in the dialogue
\item \textit{Lines to Translate} - Multiple lines of dialogue requiring translation
\item \textit{Ending} - Ending text to prompt the LLM (e.g., "Translation results:")
\end{enumerate}

Maintaining consistency in terminology translation is crucial for subtitle translation, necessitating the specific designation of terminology translations. Our general process for obtaining the terminology translation glossary used in the prompt is as follows: 1) Utilize an off-the-shelf LLM to identify and filter terminology and its translations from all dialogue; 2) Employ the identification results and an off-the-shelf LLM to train a terminology identification model, and use this model to identify and translate terminology in the test set (without ground-truth translations); 3) Retrieve the terminology appearing in the current SFT prompt's dialogue from the identification results. As terminology identification is not the primary focus of this study, we will not elaborate further on this aspect.

\subsection{Prompt Engineering Template}
\label{sec:pe}

We showcase the prompt used for the GPT-3.5, GPT-4o, and Claude 3.5 Sonnet models for the \texttt{zh}$\Rightarrow$\texttt{en} translation task (similar for other languages) in Table~\ref{tab:pe}. This prompt differ from the input to our SFT model by the inclusion of an additional translation example for LLM context learning. The prompt is structured as follows:
 
\begin{enumerate}[label=\arabic*.]
\item \textit{Preamble} - An introduction and instructions describing the task at hand
\item \textit{Example} - A few-shot example for context learning
\item \textit{Task} - Terminology information and the original dialogue to be translated, provided to the model
\item \textit{Ending} - Ending text to prompt the LLM
\end{enumerate}

\subsection{Instruction and Prompt for Quality Evaluation}
\label{sec:ioqe}

When conducting human evaluations of translation quality, it is crucial to provide evaluators with instructions that specify the evaluation perspectives, criteria, and format. This will directly influence the focus and emphasis of evaluators during the quality assessment. The provided instructions in Table~\ref{tab:instruction} are structured as follows:

\begin{enumerate}[label=\arabic*.]
\item \textit{Criteria} - Describe the professional standards that need to be followed for the translation evaluation task at hand.
\item \textit{Task} - Outline the content provided to the evaluators and the output format that must be adhered to.
\end{enumerate}

\begin{table*}[p]
\centering
\begin{tabularx}{\textwidth}{l|X}
  Preamble & \small \texttt{Please translate a series of Chinese movie/TV subtitles into English according to the following requirements: \newline 1. The translation should be colloquial, easy to understand, and consistent with the language style of the Chinese subtitles. \newline 2. Ensure that the length of the translated English subtitles matches that of the original Chinese subtitles. \newline 3. Proper nouns should be accurately translated according to the specified translations. \newline 4. When outputting, ensure that the number of translated lines matches the original text, avoid merging subtitles, and both the original text and translation must be outputted together. \newline }\\
  Terminology & \small \texttt{Translation of proper nouns: \newline \begin{CJK}{UTF8}{gkai}仙君\end{CJK} - Celestial Deity \newline \begin{CJK}{UTF8}{gkai}帝尊\end{CJK} - Your Supreme Majesty \newline \begin{CJK}{UTF8}{gkai}应渊君\end{CJK} - Sovereign Lord Yingyuan \newline ... \newline \begin{CJK}{UTF8}{gkai}玄夜\end{CJK} - Xuanye \newline \begin{CJK}{UTF8}{gkai}永夜功\end{CJK} - Eternal Darkness \newline \begin{CJK}{UTF8}{gkai}修罗尊主\end{CJK} - Asura King \newline }\\
  Lines to Translate & \small \texttt{According to the requirements previously stated, complete the following subtitle translation. \newline Original text: \newline \begin{CJK}{UTF8}{gkai}这两页为什么是空白的\end{CJK} \newline \begin{CJK}{UTF8}{gkai}有两位仙君从未更换过腰带\end{CJK} \newline \begin{CJK}{UTF8}{gkai}所以就没有记录\end{CJK} \newline \begin{CJK}{UTF8}{gkai}一位是帝尊\end{CJK} \newline \begin{CJK}{UTF8}{gkai}还有一位就是应渊君\end{CJK} \newline ... \newline \begin{CJK}{UTF8}{gkai}自你出现以后\end{CJK} \newline \begin{CJK}{UTF8}{gkai}从未更换过\end{CJK} \newline \begin{CJK}{UTF8}{gkai}应渊白日所施\end{CJK} \newline \begin{CJK}{UTF8}{gkai}竟是玄夜所创的永夜功\end{CJK} \newline \begin{CJK}{UTF8}{gkai}此功唯有修罗尊主的血脉可练\end{CJK} \newline }\\
  Ending & \small \texttt{Translation results:}
\end{tabularx}
\caption{\texttt{zh}$\Rightarrow$\texttt{en} translation prompt demonstration of SFT model.}
\label{tab:prompt}
\end{table*}

\begin{table*}[p]
\centering
\begin{tabularx}{\textwidth}{l|X}
  Response & \small \texttt{\begin{CJK}{UTF8}{gkai}这两页为什么是空白的\end{CJK}(Why are these two pages blank?) \newline \begin{CJK}{UTF8}{gkai}有两位仙君从未更换过腰带\end{CJK}(Two Celestial Deities never changed their belts.) \newline \begin{CJK}{UTF8}{gkai}所以就没有记录\end{CJK}(So, there is no record.) \newline \begin{CJK}{UTF8}{gkai}一位是帝尊\end{CJK}(His Supreme Majesty) \newline \begin{CJK}{UTF8}{gkai}还有一位就是应渊君\end{CJK}(and Sovereign Lord Yingyuan.) \newline ... \newline \begin{CJK}{UTF8}{gkai}自你出现以后\end{CJK}(Since you appeared,) \newline \begin{CJK}{UTF8}{gkai}从未更换过\end{CJK}(he had never replaced it.) \newline \begin{CJK}{UTF8}{gkai}应渊白日所施\end{CJK}(What Yingyuan executed in the daytime) \newline \begin{CJK}{UTF8}{gkai}竟是玄夜所创的永夜功\end{CJK}(was the Eternal Darkness created by Xuanye.) \newline \begin{CJK}{UTF8}{gkai}此功唯有修罗尊主的血脉可练\end{CJK}(This skill was only practiced by the bloodline of the Lord of Asura Clan.)}
\end{tabularx}
\caption{\texttt{zh}$\Rightarrow$\texttt{en} translation response demonstration of SFT model.}
\label{tab:response}
\end{table*}

\begin{table*}[p]
\centering
\begin{tabularx}{\textwidth}{l|X}
  Preamble & \small \texttt{[Requirements] \newline Please translate multiple lines of Chinese subtitles into English, adhering to the following guidelines: \newline 1. The translation should be colloquial and easily understood, maintaining consistency with the language style of the Chinese subtitles. \newline 2. Proper nouns should be translated according to the specified translations provided. \newline 3. Output the original text and translation together in the format of "Chinese original (English translation)". Ensure that subtitles are not merged, and the number of lines in the translated output matches that of the Chinese original. \newline 4. Critical requirement: The reading duration of each translated line should be consistent with the Chinese original. Ensure that the duration of the translated text is neither longer nor shorter than the original. \newline } \\
  Example & \small \texttt{[Example] \newline Proper noun translations: \newline \begin{CJK}{UTF8}{gkai}萤灯 - Yingdeng\end{CJK} \newline \begin{CJK}{UTF8}{gkai}帝君 - Your Majesty\end{CJK} \newline ... \newline \begin{CJK}{UTF8}{gkai}颜淡 - Yandan\end{CJK} \newline \begin{CJK}{UTF8}{gkai}妙法阁 - Magical Pavilion\end{CJK} \newline \newline Original text: \newline \begin{CJK}{UTF8}{gkai}萤灯姐姐\end{CJK} \newline \begin{CJK}{UTF8}{gkai}姐姐升官大喜\end{CJK} \newline ... \newline \begin{CJK}{UTF8}{gkai}甚是铺张\end{CJK} \newline \begin{CJK}{UTF8}{gkai}本君消受不起\end{CJK} \newline \newline Translation results: \newline \begin{CJK}{UTF8}{gkai}萤灯姐姐\end{CJK}(Sister Yingdeng,) \newline \begin{CJK}{UTF8}{gkai}姐姐升官大喜\end{CJK}(Congrats on your promotion,) \newline ... \newline \begin{CJK}{UTF8}{gkai}甚是铺张\end{CJK}(It's quite lavish,) \newline \begin{CJK}{UTF8}{gkai}本君消受不起\end{CJK}(I can't accept such extravagance.) \newline } \\
  Task & \small \texttt{[Task] \newline Now, following the requirements mentioned above and referring to the examples provided, translate the following Chinese dialogue into English. \newline Proper noun translations: \newline \begin{CJK}{UTF8}{gkai}有限合伙人 - limited partner\end{CJK} \newline \begin{CJK}{UTF8}{gkai}智慧社区 - smart community\end{CJK} \newline ... \newline \begin{CJK}{UTF8}{gkai}宁檬 - Ning Meng\end{CJK} \newline \begin{CJK}{UTF8}{gkai}南林股份 - Nanlin Securities\end{CJK} \newline \newline Original text: \newline \begin{CJK}{UTF8}{gkai}但是他们对于\end{CJK} \newline \begin{CJK}{UTF8}{gkai}回报的要求也非常地高\end{CJK} \newline ... \newline \begin{CJK}{UTF8}{gkai}等了这么久\end{CJK} \newline \begin{CJK}{UTF8}{gkai}终于露出马脚了\end{CJK} \newline } \\
  Ending & \small \texttt{Please directly output the translation result, ensuring to follow the format of "Chinese original (English translation)". Do not output any additional text.} \\
\end{tabularx}
\caption{\texttt{zh}$\Rightarrow$\texttt{en} translation prompt for OpenAI and Anthropic models.}
\label{tab:pe}
\end{table*}

\begin{table*}[p]
\centering
\begin{tabularx}{\textwidth}{l|X}
  Criteria & \small \texttt{\textbf{[Evaluation Criteria]} \newline \newline \textbf{1. Accuracy} \newline When assessing the accuracy of translated audiovisual dialogues, take into account the following aspects: \begin{itemize} \item \textbf{Semantic Fidelity}: Check if the original dialogue's meaning is faithfully represented in the translation and if the semantic content of the source is clearly communicated in the target language. \item \textbf{Grammatical Precision}: Evaluate the grammatical correctness of the translation, including sentence structure, verb tense, voice, and other grammatical elements. \item \textbf{Terminology Translation}: Ensure that proper nouns and specialized terms are accurately translated, preserving the original terms' semantics and context. \end{itemize} \textbf{2. Naturalness} \newline When assessing the naturalness of translated audiovisual dialogues, consider the following dimensions: \begin{itemize} \item \textbf{Coherence}: Determine if the translation reads naturally as if authored by a native speaker of the target language, and check the logical connections between sentences. \item \textbf{Readability}: Consider whether the translation is easy to read and comprehend, and if the word choice and expressions adhere to the target language conventions. \item \textbf{Fluency}: Assess whether the translation flows smoothly, has well-constructed sentences, and is free of glaring grammatical mistakes or awkward expressions. \end{itemize} \textbf{3. Vividness} \newline When assessing the vividness of translated audiovisual dialogues, review the following dimensions: \begin{itemize} \item \textbf{Stylistic Consistency}: Verify that the translation preserves the style and character traits of the original, including tonal consistency and emotional subtleties. \item \textbf{Expressiveness}: Determine if the translation captures the original dialogue's spirit and atmosphere, avoiding stiff literal translations to engage and resonate with the audience. \item \textbf{Emotion}: Check if the translation accurately reflects the characters' emotions, aligns with scene and character contexts, and emotionally connects with the target language audience. \end{itemize}}\\
  Task & \small \texttt{\textbf{[Task]} \newline \newline For each set of original dialogues, you are provided with two distinct translations (A and B). Utilize the multiple evaluation dimensions outlined in the [Evaluation Criteria] to assess the two translations for every set of original dialogues. Record your assessment results for translations A and B by marking [A is better], [B is better], or [No significant difference between A and B]. Note that your evaluation should focus on the overall quality of each set of dialogues as a whole, rather than on individual lines.}
\end{tabularx}
\caption{Guidelines for human assessment of translation quality.}
\label{tab:instruction}
\end{table*}

\end{document}